\documentclass[fleqn,usenatbib]{mnras}

\usepackage{newtxtext,newtxmath}

\usepackage[T1]{fontenc}

\DeclareRobustCommand{\VAN}[3]{#2}
\let\VANthebibliography\thebibliography
\def\thebibliography{\DeclareRobustCommand{\VAN}[3]{##3}\VANthebibliography}

\usepackage{graphicx}	
\usepackage{amsmath}	

\usepackage{bm}
\usepackage{xspace}
\newcommand{\um}{$\mu$m\xspace}
\usepackage{url}

\newcommand{\LIR}{$L_{\rm IR}$\xspace}
\newcommand{\CII}{[\ion{C}{ii}]\xspace}
\newcommand{\OIII}{[\ion{O}{iii}]\xspace}
\newcommand{\Td}{$T_{\rm d}$\xspace}

\newcommand{\Hygate}{Hygate et al. (in prep.)\xspace}
\newcommand{\Barrufet}{Barrufet et al. (in prep.)\xspace}
\newcommand{\Algera}{Algera et al. (in prep.)\xspace}
\newcommand{\Schouws}{Schouws et al. (in prep.)\xspace}
\newcommand{\Stefanon}{Stefanon et al. (in prep.)\xspace}
\newcommand{\Schneider}{Schneider et al. (in prep.)\xspace}
\newcommand{\Graziani}{Graziani et al. (in prep.)\xspace}

\newcommand{\NDetCont}{16\xspace} 

\title[REBELS: Dust Continuum Detections at $z>6.5$]{The ALMA REBELS Survey: Dust Continuum Detections at $\bm{z > 6.5}$}

\author[H. Inami et al.]{
Hanae Inami,$^{1}$\thanks{E-mail: hanae@hiroshima-u.ac.jp}
Hiddo Algera,$^{1, 2}$
Sander Schouws,$^{3}$
Laura Sommovigo,$^{4}$
Rychard Bouwens,$^{3}$
Renske Smit,$^{5}$
\newauthor
Mauro Stefanon,$^{3}$
Rebecca A. A. Bowler,$^{6}$
Ryan Endsley,$^{7}$
Andrea Ferrara,$^{4}$
Pascal Oesch,$^{8, 9}$
Daniel Stark,$^{7}$
\newauthor
Manuel Aravena,$^{10}$
Laia Barrufet,$^{8}$
Elisabete da Cunha,$^{11}$
Pratika Dayal,$^{12}$
Ilse De Looze,$^{13, 14}$
\newauthor
Yoshinobu Fudamoto,$^{15, 2}$
Valentino Gonzalez,$^{16, 17}$
Luca Graziani,$^{18, 19}$
Jacqueline A. Hodge,$^{3}$
\newauthor
Alexander P. S. Hygate,$^{3}$
Themiya Nanayakkara,$^{20}$
Andrea Pallottini,$^{4}$
Dominik A. Riechers,$^{21}$
\newauthor
Raffaella Schneider,$^{18, 22}$
Michael Topping$^{7}$
and Paul van der Werf$^{3}$
\\
$^{1}$ Hiroshima Astrophysical Science Center, Hiroshima University, 1-3-1 Kagamiyama, Higashi-Hiroshima, Hiroshima 739-8526, Japan \\
$^{2}$ National Astronomical Observatory of Japan, 2-21-1, Osawa, Mitaka, Tokyo, Japan \\
$^{3}$ Leiden Observatory, Leiden University, NL-2300 RA Leiden, Netherlands \\
$^{4}$ Scuola Normale Superiore, Piazza dei Cavalieri 7, 56126 Pisa, Italy \\
$^{5}$ Astrophysics Research Institute, Liverpool John Moores University, 146 Brownlow Hill, Liverpool L3 5RF, United Kingdom \\
$^{6}$ Jodrell Bank Centre for Astrophysics, Department of Physics and Astronomy, School of Natural Sciences, The University of Manchester, Manchester, M13 9PL, UK
\\
$^{7}$ Steward Observatory, University of Arizona, 933 N Cherry Ave, Tucson, AZ 85721, United States \\
$^{8}$ Observatoire de Gen\'eve, 1290 Versoix, Switzerland \\
$^{9}$ Cosmic Dawn Center (DAWN), Niels Bohr Institute, University of Copenhagen, Jagtvej 128, K{\o}benhavn N, DK-2200, Denmark \\
$^{10}$ Nucleo de Astronomia, Facultad de Ingenieria y Ciencias, Universidad Diego Portales, Av. Ejercito 441, Santiago, Chile \\
$^{11}$ International Centre for Radio Astronomy Research, Uni- versity of Western Australia, 35 Stirling Hwy, Crawley,26WA 6009, Australia \\
$^{12}$ Kapteyn Astronomical Institute, University of Groningen, P.O. Box 800, 9700 AV Groningen, The Netherlands \\
$^{13}$ Sterrenkundig Observatorium, Ghent University, Krijgslaan 281 - S9, 9000 Gent, Belgium \\
$^{14}$ Dept. of Physics \& Astronomy, University College London, Gower Street, London WC1E 6BT, United Kingdom \\
$^{15}$ Research Institute for Science and Engineering, Waseda University, 3-4-1 Okubo, Shinjuku, Tokyo 169-8555, Japan \\
$^{16}$ Departmento de Astronomia, Universidad de Chile, Casilla 36-D, Santiago 7591245, Chile \\
$^{17}$ Centro de Astrofisica y Tecnologias Afines (CATA), Camino del Observatorio 1515, Las Condes, Santiago, 7591245, Chile \\
$^{18}$ Dipartimento di Fisica, Sapienza, Universita di Roma, Pi- azzale Aldo Moro 5, I-00185 Roma, Italy \\
$^{19}$ INAF/Osservatorio Astrofisico di Arcetri, Largo E. Femi 5, I-50125 Firenze, Italy \\
$^{20}$ Centre for Astrophysics \& Supercomputing, Swinburne University of Technology, PO Box 218, Hawthorn, VIC 3112, Australia \\
$^{21}$ I. Physikalisches Institut, Universit\"at zu K\"oln, Z\"ulpicher Strasse 77, D-50937 K\"oln, Germany \\
$^{22}$ INAF/Osservatorio Astronomico di Roma, via Frascati 33, 00078 Monte Porzio Catone, Roma, Italy
}

\date{Accepted 2022 June 23. Received 2022 May 23; in original form 2022 March 11}

\pubyear{2015}

\begin{document}
\label{firstpage}
\pagerange{\pageref{firstpage}--\pageref{lastpage}}
\maketitle

\begin{abstract} 
 We report 18 dust continuum detections ($\geq 3.3\sigma$) 
   at $\sim88$\um and 158\um
   out of 49 ultraviolet(UV)-bright galaxies 
  ($M_{\rm UV} < -21.3$\,mag)
 at $z>6.5$, observed by the Cycle-7 ALMA Large
   Program, REBELS and its pilot programs. 
   This has more than tripled the number of
   dust continuum detections known at $z>6.5$. Out of these 18
   detections, 12 are reported for the first time as part of REBELS.
  In addition, 15 of the dust continuum detected 
  galaxies also show a
    [\ion{C}{ii}]$_{\rm 158{\rm \mu m}}$ emission line, providing us
    with accurate redshifts.  We anticipate more
  line emission detections from six targets (including three 
  continuum detected targets) where observations are still ongoing.
  We estimate that all of the sources have an infrared (IR) luminosity
  ($L_{\rm IR}$) in a range of $3-8 \times 10^{11}\,{\rm L_\odot}$, except
  for one with $L_{\rm IR} = 1.5^{+0.8}_{-0.5} \times 10^{12}\,
  \,{\rm L_{\odot}}$.
  Their fraction of obscured star formation is significant at 
  $\gtrsim 50\%$, despite being UV-selected galaxies.  
    Some of the dust continuum
  detected galaxies show spatial offsets ($\sim 0.5-1.5\arcsec$)
  between the rest-UV and far-IR emission peaks.
  These separations could imply spatially decoupled 
  phases of obscured and unobscured star formation, but a higher spatial
  resolution observation is required to confirm this.
  REBELS offers the best available statistical constraints on obscured
  star formation in UV-luminous galaxies at $z > 6.5$.
\end{abstract}




\section{Introduction} \label{sec:intro}

While cosmic dust constitutes only a small fraction ($< 1\%$) of the
total mass of the interstellar medium (ISM), it plays a major role in
characterising galaxy properties and in influencing
  galaxy evolution \citep[][]{Spar00}.  Dust absorbs and scatters
ultraviolet (UV) and optical emission, and reradiates 
the absorbed energy in the infrared
(IR) as thermal emission.  Thus, far-infrared emission from dust
traces obscured star formation, which is key to
  obtaining a complete census of star formation in the Universe.  The
infrared star formation rate density increases with redshift at least
up to $z\sim2-3$ and dominates the total star formation rate density
in this redshift range \citep[e.g.,][]{Mada14,Magn13,Grup13,Nova17}.

With the advent of the Atacama Large Millimetre/submillimetre Array
(ALMA), detections of dusty star-forming galaxies at higher redshift
($z \gtrsim 3$) have been steadily increasing, enabling some more
complete studies of dust embedded star formation in the distant
universe \citep[for a review see][]{Hodg20}. In particular, large
surveys have been pushing forwards our understanding of obscured star
formation at high redshifts. Multiple deep surveys in the GOODS-South
field have characterised the detailed properties of galaxies, mostly at
$z=2-4$, that contribute to obscured star formation \citep{Dunl17,
  Fran20a, Fran20b, Yama19, Yama20, Gonz20, Arav20}. 
A prediction of obscured star formation out
to $z=7$ has been made by \cite{Zava21} (see also \citealt{Case21}),
combining a shallower but wide area 2mm survey with the ultra-deep 1mm
and 3mm surveys \citep{Gonz19,Gonz20}.  The ALMA Large Program to
INvestigate \CII at Early times (ALPINE) has revealed dusty star
formation at $z\sim5.5$ with their main target galaxies \citep{Khus21}
and at $1 \lesssim z \lesssim 6$ with serendipitously detected
galaxies in the fields that they targeted \citep{Grup20}.  The current
census seems to indicate that the obscured star formation density
starts to decrease from $z\sim2-3$ towards higher redshift, but how
fast it declines is still uncertain.

In addition, direct comparisons between the far-IR and UV emission
have revealed that there is potentially an evolution in the
relationship between the infrared excess (${\rm IRX} = L_{\rm IR}/L_{\rm
  UV}$) and the UV spectral slope ($\beta_{\rm UV}$) or stellar mass
(${\rm M_{*}}$) \citep{Capa15, Bouw16, Bari17, McLu18, Kopr18, Bouw20,
  Fuda20b, Bowl21b}.  Beside uncertainties on $L_{\rm
    IR}$ often being large at high redshift, this could imply that a
different attenuation curve may need to be adopted when UV and dust
emission co-exist.  However, there are some cases that show spatial
offsets between UV and dust emission, making it harder to interpret
the difference seen in IRX-$\beta_{\rm UV}$ (or ${\rm M_{*}}$) in the
distant (and local) universe
\citep[e.g.,][]{Hodg12,Hodg16,Kopr16,Carn17,Lapo17a,Popp17c,Howe10}.
Furthermore, if a spatial offset between UV and far-IR is common among
high redshift galaxies \citep[e.g.,][]{Behr18,Lian19,Ma19,Somm20},
stacking of far-IR data based on UV locations will not present the
full picture of the dust emission.

Dust is not only responsible for obscured star formation, but also
regulates the evolution of galaxies by, for example, contributing to
heating of the gas in the ISM by photoelectric heating
\citep[e.g.,][]{Nath99} and facilitating molecular gas ($\rm H_2$)
formation on its grain surfaces \citep[e.g.,][]{Wake17}. However, the
first emergence of dust and its buildup at the beginning of the
universe is still unknown.  Attempts at searching for dusty galaxies
in the distant universe have been extended to the epoch of
reionisation.  Recent ALMA observations, despite the small number of
detections, enable us to discern mounting evidence that significant
obscured star formation was already taking place in some normal
galaxies as early as $\sim 600$\,Myr after the Big Bang \citep{Wats15,
  Knud17, Lapo17a, Bowl18, Hash19a, Suga21, Tamu19, Bakx20, Bakx21,
  Scho21, Bowl21b, Lapo21a, Fuda21}.  A large number
  of theoretical studies have also been conducted to try to guide
  interpretations of these observed galaxies
  \citep[e.g.,][]{Aoya17,Aoya18,Gino18,Behr18,GrazL20,Pall22}.

A mechanism of fast dust formation or grain growth
  may be required at such an early era of the universe ($\lesssim
  1$\,Gyr) to produce the observed dust mass 
  \citep{Manc15,Manc16,Gino18,GrazL20}.
On the other hand, there are other models coupling the baryonic and 
dust assembly of early galaxies, which show that even 
invoking ISM dust growth timescales as short as 0.3\,Myr can only 
scale up the dust mass by a factor of 2 \citep{Daya22}. 
For low mass stars ($< 8\,{\rm M_{\odot}}$), their lifetime is too long 
 to evolve into the AGB phase by $z \sim 7$. Thus, supernovae are likely 
 the major stellar source contributor producing the dust in
galaxies at $z \gtrsim 7$ \citep[][]{Lesn19,Manc15,Somm20,Daya22}.   
With existing observations, we have increasingly gained some perspective on
the dust properties of some galaxies at high redshifts, but the
limited number of detections has made it difficult to characterise the
properties in a statistical way.

The ALMA Cycle-7 Large Program, Reionization-Era Bright Emission Line
Survey \citep[REBELS,][]{Bouw21b}, aims to obtain a statistical sample
of luminous star-forming galaxies at $z > 6.5$. 
REBELS is designed to acquire spectroscopic redshifts from an ISM cooling 
line and simultaneously detect dust continuum emission.  This on-going large
survey targets 40 UV-luminous Lyman-break galaxies selected 
over $\sim7\,\deg^2$ to search for the [\ion{C}{ii}]$_{158\,{\rm \mu m}}$ or
[\ion{O}{iii}]$_{88\,{\rm \mu m}}$ emission lines with spectral scans.
As shown in \cite{Bouw21b} and the reports from the pilot programs
\citep{Smit18,Scho21,Scho22}, the observational strategy has proven to
dramatically increase the number of spectroscopically confirmed
galaxies and dust continuum detections at $z > 6.5$. A larger sample
allows us to explore the prevalence of dust-rich galaxies,
characterise their dust properties, and study dust buildup in the
early epoch of the universe (e.g., \citealt{Daya22}, \citealt{Ferr22},
\citealt{Somm22}, \Schneider, \Graziani).  The REBELS
  survey has also proved its power by making serendipitous detections
  at $z\sim7$ of two heavily obscured star-forming galaxies without an
  optical counterpart \citep{Fuda21}.  The detections of \CII
emission in REBELS are reported by \Schouws.

In the current paper, we present dust continuum detections and flux
extractions of the UV-selected luminous galaxies 
at $z > 6.5$ from the REBELS and pilot programs. Together with 
the dust properties, we also
show galaxy global properties derived from the UV emission (\Stefanon)
of the dust continuum detected galaxies. In addition, we compare the
dust morphologies of these galaxies to their UV morphologies.

The paper is organised as follows. In \S2, we describe the REBELS
observations and the data reduction, followed by presenting the
ancillary data used in this work. Then in \S3, we present the method
to identify dust continuum emission, measure the dust continuum flux,
and calculate the total infrared luminosity. Galaxy properties of the
dust continuum detected sources are shown in \S4. We discuss
dust morphologies and spatial offsets between the
rest-UV and IR emission in \S5. 
Lastly, we summarise our findings in \S6.

We adopt a flat $\Lambda$CDM cosmology with $H=70\,{\rm
  km\,s^{-1}\,Mpc^{-1}}$, $\Omega_{\rm M}=0.3$ and $\Omega_{\rm
  \Lambda}=0.7$. All magnitudes are expressed in the AB system
\citep{Oke83}. For star formation rates (SFRs) and
  stellar masses ($M_*$), we adopt a Chabrier initial mass function
  \citep[IMF,][]{Chab03}.

\begin{figure*}
  \includegraphics[width=\textwidth]{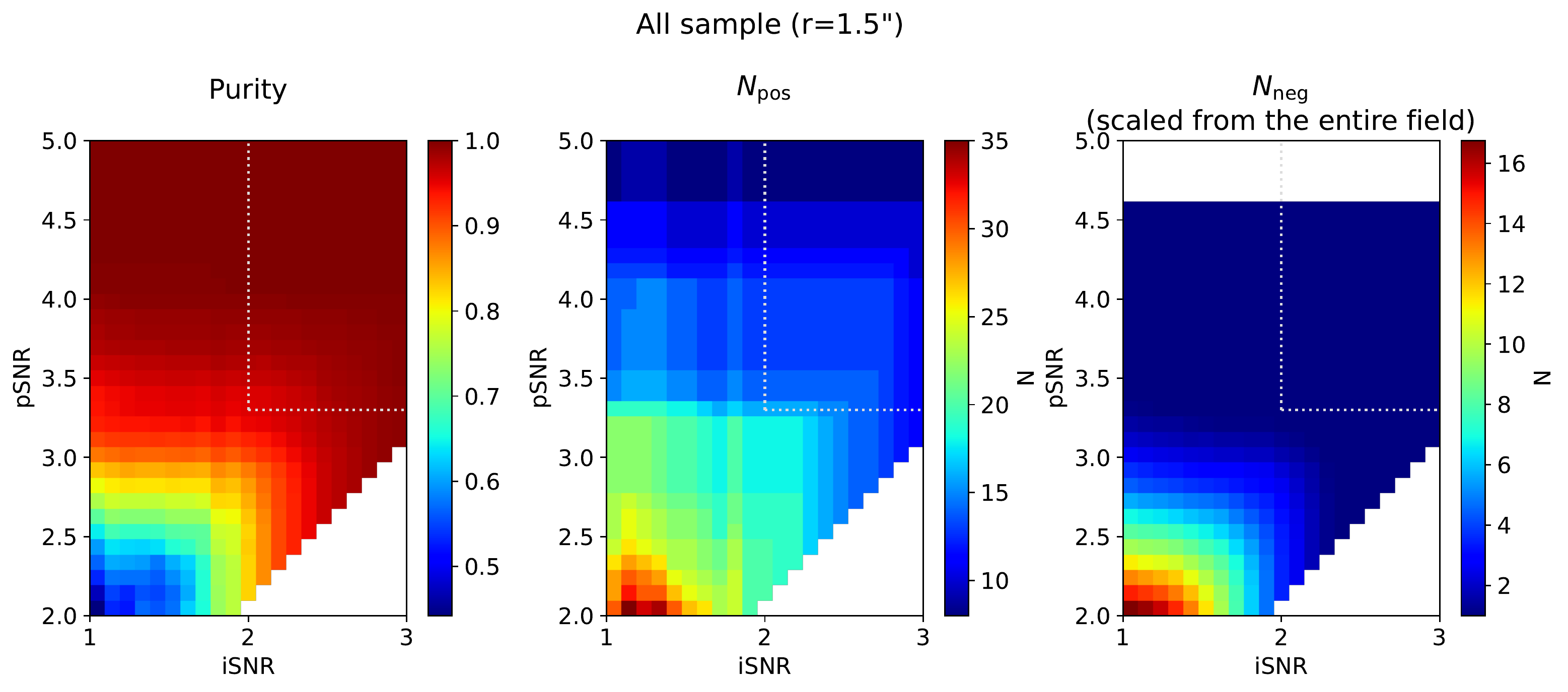}
  \caption{ From left to right, purity ($p$), the number of detections
    in the science (positive) image ($N_{\rm pos}$), and the number of
    detections in the reversed science (negative) image ($N_{\rm
      neg}$). The detections in all of the 40 pointing fields have been combined.
    The parameters iSNR and pSNR are used
      in \texttt{PyBDSF} to determine the boundary of an island and
      the peak of a source, respectively. 
     The detection search for $N_{\rm pos}$ was performed 
     around the phase centre with a 1.5\arcsec radius, 
     while $N_{\rm neg}$ was scaled down to the same area from 
     the detection search of the entire pointing field. 
    The white area in
      the bottom right of each panel (${\rm iSNR} > {\rm pSNR}$) was
      excluded from the analysis.    The dotted box
    indicates the thresholds (pSNR, iSNR)\,$= (3.3, 2.0)$ for reliable
    detections ($p\geq95\%$).
    \label{fig:purity_2D}
  }
\end{figure*}

\section{Observations and Data Reduction} \label{sec:reduction}

\subsection{The REBELS program} \label{subsec:dust_img}

REBELS is an ALMA Large Program in Cycle-7 (2019.1.01634.L, PI:
Bouwens) that systematically targets 40 UV-luminous (${\rm M_{UV}} <
-21.3$\,mag) star-forming galaxies in the Epoch of Reionization ($z >
6.5$) to perform a spectral scan for the [\ion{C}{ii}]$_{\rm {158\mu
    m}}$ or [\ion{O}{iii}]$_{\rm {88\mu m}}$ emission line.
The galaxies targeted by the REBELS survey were searched for in a
total area of $\sim 7\,{\rm deg}^2$, including COSMOS/UltraVISTA
\citep{Scov07, McCr12}, UKIDSS/UDS and VIDEO/XMM-LSS \citep{Jarv13,
  Lawr07}, CANDELS \citep{Grog11}, CLASH \citep{Post12}, and
BoRG/HIPPIES pure-parallel \citep{Tren11, Yan11}.  In these search
fields, our targets were selected to have a UV SFR of $> 10 \,
{{\rm M_{\odot}} \, {\rm yr}^{-1}}$, a strong Lyman break, and 
a photometric redshift of $6.5 \lesssim z_{\rm phot} \lesssim 9.5$.

The survey design is detailed in \cite{Bouw21b} and here we provide a
brief summary.  To maximise the spectral scan efficiency, for the
targets with $z_{\rm phot} < 8.5$ we have observed the \CII line with
Band~5 or 6. For the targets at higher redshift, the \OIII line
has been observed with Band~7.
Dust continuum emission is simultaneously observed during the spectral
scans of the \CII or \OIII emission line.
The observations were carried out with the most compact array
configuration (C43-1 and C43-2) with a synthesised beam of $\sim
1.2-1.6\arcsec$ full width at half maximum (FWHM).

Most of the observations were performed between 2019/11 and 2020/01.
There are still six targets whose observations are to be completed
(REBELS-04, 06, 11, 16, 24, and 37). The completed 
observations constitute 87\% (60.6\,hours) of the approved total 
observing time of 70\,hours.  
In addition, we include the results from our pilot observations 
in this paper. In our first pilot study we detected \CII in two galaxies 
at $z \sim 6.8$ \citep{Smit18} and in our second pilot study we detected 
dust continuum in six $z \sim 7-8$ galaxies \citep{Scho21} and \CII in
three $z \sim 7$ galaxies \citep{Scho22} (see also Table~\ref{tbl:det_src}). 
For six of the galaxies from the pilot observations in which the \CII line 
was not detected, they have been targeted in the main REBELS survey 
to scan the remainder of the redshift probability distribution. 
The observational setups for the pilot and REBELS 
observations are the same. We therefore include the visibility 
measurement sets from the pilot observations for those six sources 
when creating the dust continuum maps to maximise sensitivity.
In total, we have targeted 49 UV-selected galaxies in the REBELS main and 
pilot surveys at $z > 6.5$.

\subsection{ALMA Data Reduction and Dust Continuum Images} \label{subsec:data}

The data reduction and calibration were carried out with the
observatory-provided standard pipeline using the Common Astronomy
Software Application (\textsc{CASA}) package \citep[version
  5.6.1;][]{McMu07}. The datacubes were then imaged using the
\textsc{tclean} task with automasking to clean down to $2\sigma$ with 
a cell pixel size of $0.15\arcsec$ and natural weighting. 
The dust continuum maps were created using the calibrated datacubes by
excluding the frequency range that covers $\rm 2 \times FWHM$ around a
detected emission line, \CII or \OIII, of the targeted source in the
phase centre.  A more detailed description, in particular for 
the procedures related to identifying an emission line in the data, 
can be found in \Schouws.

\subsection{SED fittings of UV-NIR photometries} \label{subsec:sed}

In this work, we use the optical and near-IR
  (NIR) images taken by the surveys listed in
  \S\ref{subsec:dust_img}. Based on these data, we investigate
rest-frame UV properties of the dust continuum detected sources in our
sample. The physical properties of the REBELS galaxies were derived
from rest-UV photometries and SED fittings using the \textsc{BEAGLE}
code \citep{Chev16}.  In the current paper, we did
  not include any dust continuum measurements (or upper limits) in the
  \textsc{BEAGLE} SED fits.  Here we briefly summarise the
procedure. For the full description of the SED
  fitting as well as the available optical and near-IR (NIR) data, 
  we refer to \Stefanon.

\textsc{BEAGLE} utilises the synthetic models of stellar and nebular
emission from \cite{Gutk16} which combine the latest stellar
population synthesis models of \cite{BC03} with the photoionisation
models from \textsc{CLOUDY} \citep{Ferl13}. We assumed a constant star
formation history (SFH)~\footnote{In the paper, 
we adopt the constant SFH to facilitate comparisons to earlier studies. 
Note that non-parametric treatments of SFH for galaxies during 
the epoch of reionisation could lead to a higher stellar mass.
We refer to \cite{Topp22} for the adoption of non-parametric SFHs of 
the REBELS galaxies and comparisons to the constant SFH values.}, 
a sub-solar metallicity \citep[$0.2 Z_{\odot}$; motivated by][]{Star17,DeBa19}, 
and a \cite{Chab03} IMF mass cutoff of $0.1-300,{\rm M_{\odot}}$ 
to reduce a bias towards extremely young stellar population ages.  
We adopted the \cite{Calz00} dust extinction curve. Based on these
  assumptions, we derived UV magnitude ($M_{\rm UV}$), $\beta_{\rm
    UV}$, dust extinction in the $V$-band ($A_V$), stellar age, and
  the equivalent width of the sum of the [\ion{O}{iii}] and H$\beta$ 
  emission lines via SED fitting.

\subsection{Astrometry Correction of the NIR images} \label{subsec:astrometry}

We also explore spatial offsets between rest-UV (NIR
in the observed frame) and dust emission. For all of the REBELS
sample, there are $J$, $H$, and $K$-band images available, 
except for the two galaxies selected from the data taken with 
the {\it Hubble} Space Telescope, REBELS-16 (MACS0429-Z1) and 
REBELS-40 (Super8-1), which do not have a $K$-band image.

The spatial offset analysis requires that all astrometry is well
aligned.  Based on the beam size at $\sim1$mm, we find that the
expected astrometric uncertainty of the ALMA observations is $\sim
0.2\arcsec$ and $\sim 0.5\arcsec$ at the signal-to-noise ratios (SNR)
of 5 and 3, respectively, for our continuum detections.  We did not
find any counterparts for either the main REBELS targets
  or serendipitous sources in the pointing fields with solid
detections or any matching point sources in public 
catalogues of the Very Large Array \citep[VLA; e.g.,][]{Smol17} 
to perform a further examination for the astrometric alignment.

For the rest-UV data taken in the NIR bands, we either used the
public images that were already aligned to a Gaia catalogue (UltraVISTA)
or applied the astrometric correction using the Gaia Data Release 3
catalogue for the XMM and the {\it HST} fields \citep{Gaia16,Gaia21} with the
procedure as follows.  For the REBELS sources in the XMM fields (XMM1,
2 or 3, which consists of three separate tiles according to the VISTA
VIRCAM footprint), the correction was not applied to an entire mosaic
image but rather to a $6\arcmin \times 6\arcmin$ cropped image with a
REBELS source at the centre to improve the local
  astrometry. An exception is REBELS-08; because it lies near the
edge of the original $JHK$ images, we shifted the centre of the
cropped image by $-1.8\arcmin$. For the two sources (REBELS-16 and 40)
observed with the {\it HST}, we use the full WFC3
image to maximise the number of Gaia sources available for
registration.

We then crossmatched the Gaia sources in the NIR images to
calculate their offsets in the XY directions. The
  centres of the matched sources in the NIR images were found by
  Gaussian fitting.
The Gaia sources with a large parallax angle ($> 10$mas) were not used
for the astrometric alignment. Proper motion was corrected using the
Gaia reference epoch and the date when a NIR image was taken. When
the NIR data were taken over a certain period of time (e.g.,
mosaic), the mid-point of the first and last date was used.  The
correction factor was calculated using the $3\sigma$-clipped mean
separation between the Gaia sources and the same sources found in the
NIR image.  Although each NIR band image has its own
correction factor, we used their mean value to correct for all
NIR images taken within each observing program.
The difference between the individual offsets and the
  mean offsets are small (typically a few milliarcseconds).
This correction improved the astrometry to an RMS of
$\approx 30-120$mas between the Gaia stars and the corresponding
sources in the NIR image.

\begin{center}
  \begin{center}
\begin{table*}
\caption{Summary of the dust continuum flux measurements of the REBELS targets (the first 40 sources) and the pilot targets (the last nine sources). For the sources with a detection, we quote their Gaussian fit (\texttt{CASA/imfit}) results, while for a non detection, the peak emission upper limit ($3\sigma$) is shown. The reported flux for REBELS-25 is the integrated flux, but the peak fluxes for the rest of the sources (\S\ref{subsec:flux_meas}). The rest-frame 158\um continuum flux density is not corrected for the CMB effect, while the correction has been applied to $L_{\rm IR}$. The redshifts are spectroscopic redshifts for the sources with a \CII detection, otherwise they are photometric redshifts indicated by a dagger. The flux density from literature (uncorrected for the CMB effect) is taken from \protect\cite{Scho22}, \protect\cite{Smit18}, and \protect\cite{Bowl21b}, indicated as S22, S18, and B21, respectively. The two sources with the asterisk from the pilot program, UVISTA-Y-008 and UVISTA-Y-010, are listed as UVISTA-Y-007 and UVISTA-Y-009, respectively, in \protect\cite{Scho22}. The ALMA source names shown in parentheses are the ones used in the ALMA Science Archive. A machine readable table is available at \url{http://vizier.u-strasbg.fr/viz-bin/VizieR?-source=J/MNRAS/515/3126} on CDS.
\label{tbl:det_src}}
\begin{tabular}{lllccccccc}
  \hline
  REBELS ID & ALMA source name &    Redshift & Observed frequency &      Continuum flux &  SNR &             $L_{\rm IR}$ & $F_{\nu}$ from literature & Literature \\ 
            &                  &             &                GHz &   $\mathrm{\mu Jy}$ &      &       $10^{11}L_{\odot}$ &         $\mathrm{\mu Jy}$ &            \\ 
  \hline
  REBELS-01 & XMM1-Z-276466    & $   7.1771$ &             228.12 & $<              50$ &  ... & $<                  2.9$ &                       ... & ...        \\ 
  REBELS-02 & XMM1-35779       & $6.64^\dag$ &             248.69 & $<              49$ &  ... & $<                  2.5$ &                       ... & ...        \\ 
  REBELS-03 & XMM1-Z-1664      & $   6.9695$ &             238.57 & $<              51$ &  ... & $<                  2.8$ &                       ... & ...        \\ 
  REBELS-04 & XMM-J-355        & $8.57^\dag$ &             347.77 & $    68 \pm     20$ &  3.4 & $  4.5\,^{+ 1.6}_{-2.7}$ &                       ... & ...        \\ 
  REBELS-05 & XMM1-1591        & $   6.4963$ &             248.69 & $    67 \pm     13$ &  5.3 & $  3.2\,^{+ 1.9}_{-1.2}$ &                       ... & ...        \\ 
  REBELS-06 & XMM1-Z-151269    & $6.80^\dag$ &             244.00 & $    77 \pm     15$ &  5.0 & $  4.0\,^{+ 2.3}_{-1.5}$ &                       ... & ...        \\ 
  REBELS-07 & XMM1-Z-1510      & $7.15^\dag$ &             230.97 & $<              49$ &  ... & $<                  2.8$ &                       ... & ...        \\ 
  REBELS-08 & XMM1-67420       & $   6.7495$ &             246.24 & $   101 \pm     20$ &  5.1 & $  5.2\,^{+ 3.0}_{-2.0}$ &                       ... & ...        \\ 
  REBELS-09 & XMM2-Z-1116453   & $7.61^\dag$ &             223.26 & $<              53$ &  ... & $<                  3.4$ &                       ... & ...        \\ 
  REBELS-10 & XMM2-Z-564239    & $7.42^\dag$ &             223.26 & $<              56$ &  ... & $<                  3.4$ &                       ... & ...        \\ 
  REBELS-11 & XMM3-Y-217016    & $8.24^\dag$ &             347.77 & $<              93$ &  ... & $<                  5.8$ &                       ... & ...        \\ 
  REBELS-12 & XMM3-Z-110958    & $   7.3459$ &             228.12 & $    87 \pm     24$ &  3.6 & $  5.2\,^{+ 3.1}_{-2.2}$ &                       ... & ...        \\ 
            &                  &             &                    & ($   48 \pm     13$ & 3.8) &                      ... &                       ... & ...        \\ 
            &                  &             &                    & ($   39 \pm     12$ & 3.3) &                      ... &                       ... & ...        \\ 
  REBELS-13 & XMM-J-6787       & $8.19^\dag$ &             349.67 & $<              87$ &  ... & $<                  5.4$ &                       ... & ...        \\ 
  REBELS-14 & XMM3-Z-432815    & $   7.0842$ &             238.57 & $    60 \pm     15$ &  4.1 & $  3.3\,^{+ 2.0}_{-1.4}$ &                       ... & ...        \\ 
  REBELS-15 & XMM3-Z-1122596   & $   6.8752$ &             244.00 & $<              68$ &  ... & $<                  3.6$ &                       ... & ...        \\ 
  REBELS-16 & MACS0429-Z1      & $6.70^\dag$ &             245.90 & $<              71$ &  ... & $<                  3.6$ &                       ... & ...        \\ 
  REBELS-17 & UVISTA-Z-1373    & $   6.5376$ &             248.12 & $<              80$ &  ... & $<                  3.9$ &                       ... & ...        \\ 
  REBELS-18 & UVISTA-Y-001     & $   7.6750$ &             204.72 & $    53 \pm     10$ &  5.3 & $  3.5\,^{+ 2.0}_{-1.3}$ & $     73.0 \pm      20.0$ & S22        \\ 
  REBELS-19 & UVISTA-Y-879     & $   7.3701$ &             227.42 & $    71 \pm     20$ &  3.5 & $  4.3\,^{+ 2.6}_{-1.9}$ &                       ... & ...        \\ 
            &                  &             &                    & ($   32 \pm     11$ & 3.1) &                      ... &                       ... & ...        \\ 
            &                  &             &                    & ($   39 \pm     10$ & 3.9) &                      ... &                       ... & ...        \\ 
  REBELS-20 & UVISTA-Z-734     & $7.12^\dag$ &             233.79 & $<              76$ &  ... & $<                  4.3$ &                       ... & ...        \\ 
  REBELS-21 & UVISTA-Z-013     & $6.59^\dag$ &             240.75 & $<              53$ &  ... & $<                  2.6$ & $<                  45.0$ & S22        \\ 
  REBELS-22 & UVISTA-Y-657     & $7.48^\dag$ &             227.42 & $<              46$ &  ... & $<                  2.8$ &                       ... & ...        \\ 
  REBELS-23 & UVISTA-Z-1410    & $   6.6447$ &             246.96 & $<              78$ &  ... & $<                  3.9$ &                       ... & ...        \\ 
  REBELS-24 & UVISTA-Y-005     & $8.35^\dag$ &             202.03 & $<              42$ &  ... & $<                  3.2$ & $<                  38.7$ & S22        \\ 
  REBELS-25 & UVISTA-Y-003     & $   7.3065$ &             227.28 & $   260 \pm     22$ & 11.7 & $ 15.4\,^{+ 8.4}_{-5.2}$ & $    241.0 \pm      30.0$ & S22        \\ 
  REBELS-26 & UVISTA-Z-011     & $   6.5981$ &             246.57 & $<              95$ &  ... & $<                  4.7$ &                       ... & ...        \\ 
  REBELS-27 & UVISTA-Y-004     & $   7.0898$ &             227.28 & $    51 \pm     10$ &  5.1 & $  2.9\,^{+ 1.6}_{-1.1}$ & $     65.0 \pm      17.0$ & S22        \\ 
  REBELS-28 & UVISTA-Z-1595    & $   6.9433$ &             244.66 & $<              65$ &  ... & $<                  3.5$ &                       ... & ...        \\ 
  REBELS-29 & UVISTA-Z-004     & $   6.6847$ &             240.42 & $    56 \pm     13$ &  4.4 & $  2.9\,^{+ 1.7}_{-1.1}$ & $     50.0 \pm      15.0$ & B21        \\ 
  REBELS-30 & UVISTA-Z-009     & $   6.9823$ &             242.41 & $<              54$ &  ... & $<                  3.0$ & $     81.7 \pm      48.6$ & B21        \\ 
  REBELS-31 & UVISTA-Z-005     & $6.68^\dag$ &             248.12 & $<              84$ &  ... & $<                  4.2$ & $<                  21.5$ & B21        \\ 
  REBELS-32 & UVISTA-Z-049     & $   6.7290$ &             244.54 & $    60 \pm     17$ &  3.5 & $  3.1\,^{+ 1.9}_{-1.3}$ &                       ... & ...        \\ 
  REBELS-33 & UVISTA-Z-018     & $6.67^\dag$ &             248.12 & $<              80$ &  ... & $<                  4.0$ &                       ... & ...        \\ 
  REBELS-34 & UVISTA-Z-002     & $   6.6335$ &             248.12 & $<              75$ &  ... & $<                  3.8$ & $     53.2 \pm      32.5$ & B21        \\ 
  REBELS-35 & UVISTA-Z-003     & $6.97^\dag$ &             239.17 & $<              70$ &  ... & $<                  3.8$ &                       ... & ...        \\ 
  REBELS-36 & UVISTA-Y-002     & $   7.6772$ &             208.49 & $<              42$ &  ... & $<                  2.7$ & $<                  39.9$ & S22        \\ 
  REBELS-37 & UVISTA-J-1212    & $7.75^\dag$ &             349.66 & $    97 \pm     16$ &  6.1 & $  5.6\,^{+ 1.4}_{-3.1}$ &                       ... & ...        \\ 
  REBELS-38 & UVISTA-Z-349     & $   6.5770$ &             246.57 & $   163 \pm     23$ &  7.1 & $  8.0\,^{+ 4.5}_{-2.9}$ &                       ... & ...        \\ 
  REBELS-39 & UVISTA-Z-068     & $   6.8449$ &             244.54 & $    80 \pm     16$ &  4.9 & $  4.2\,^{+ 2.4}_{-1.6}$ &                       ... & ...        \\ 
  REBELS-40 & Super8-1         & $   7.3650$ &             223.26 & $    48 \pm     13$ &  3.7 & $  2.9\,^{+ 1.7}_{-1.2}$ &                       ... & ...        \\ 
  \hline
  REBELS-P1 & UVISTA-Z-007     & $   6.7496$ &             245.65 & $<              67$ &  ... & $<                  3.4$ & $<                  52.2$ & S22        \\ 
  REBELS-P2 & UVISTA-Y-008$^*$ (Y8) & $8.47^\dag$ &             230.36 & $<              65$ &  ... & $<                  4.9$ & $<                  53.7$ & S22        \\ 
  REBELS-P3 & UVISTA-Y-010$^*$ (Y10) & $7.69^\dag$ &             230.36 & $<              66$ &  ... & $<                  4.3$ & $<                  53.6$ & S22        \\ 
  REBELS-P4 & UVISTA-Y-006 (Y6) & $8.32^\dag$ &             203.08 & $<              53$ &  ... & $<                  4.0$ & $<                  39.6$ & S22        \\ 
  REBELS-P5 & UVISTA-Z-010     & $7.19^\dag$ &             233.99 & $<              59$ &  ... & $<                  3.4$ & $<                  44.1$ & S22        \\ 
  REBELS-P6 & COS-2987030247   & $   6.8075$ &             243.00 & $<             109$ &  ... & $<                  5.7$ & $<                  75.0$ & S18        \\ 
  REBELS-P7 & UVISTA-Z-019     & $   6.7534$ &             245.65 & $    61 \pm     16$ &  3.8 & $  3.2\,^{+ 1.9}_{-1.3}$ & $     66.0 \pm      23.0$ & S22        \\ 
  REBELS-P8 & COS-3018555981   & $   6.8537$ &             243.00 & $<             122$ &  ... & $<                  6.4$ & $     65.0 \pm      13.0$ & S22        \\ 
  REBELS-P9 & UVISTA-Z-001     & $   7.0599$ &             233.99 & $    55 \pm     17$ &  3.3 & $  3.1\,^{+ 1.9}_{-1.4}$ & $     48.1 \pm      19.6$, $    104.0 \pm      43.0$ & B21,S22    \\ 
  \hline
\end{tabular}
\end{table*}
\end{center}

\end{center}

\begin{figure}
  \includegraphics[width=\columnwidth, trim=0 0 0 40, clip]{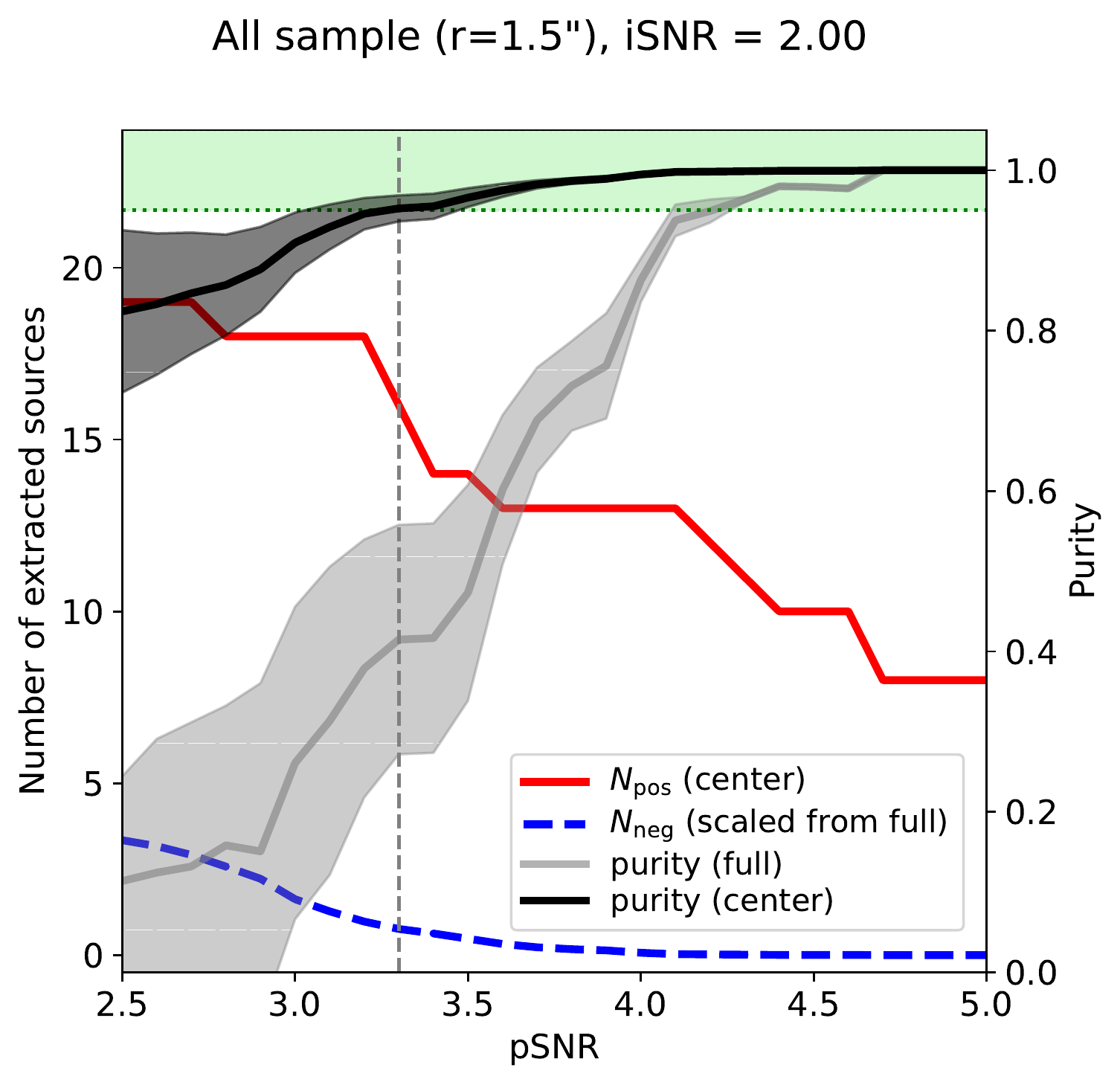}
  \caption{The numbers of positive ($N_{\rm pos}$; the red solid line)
    and negative ($N_{\rm neg}$; the blue dashed line) detections (the
    left axis) and purity (the right axis) as a function of pSNR
    when iSNR is fixed at 2.0. The dark and light grey colour lines
    indicate the purity of the central $r=1.5\arcsec$ and the full
    pointing field, respectively. The shaded areas represent their
    standard deviations.  At pSNR\,$=3.3$ (the vertical dashed line),
    purity of $\geq95\%$ (the light green area) is achieved for the REBELS main
    targets that are expected to be detected within the central
    $r=1.5\arcsec$ of the phase centre.
    \label{fig:purity_1D}
  }
\end{figure}

\begin{figure*}
  \includegraphics[width=\textwidth, clip, trim=0 0 0 30]{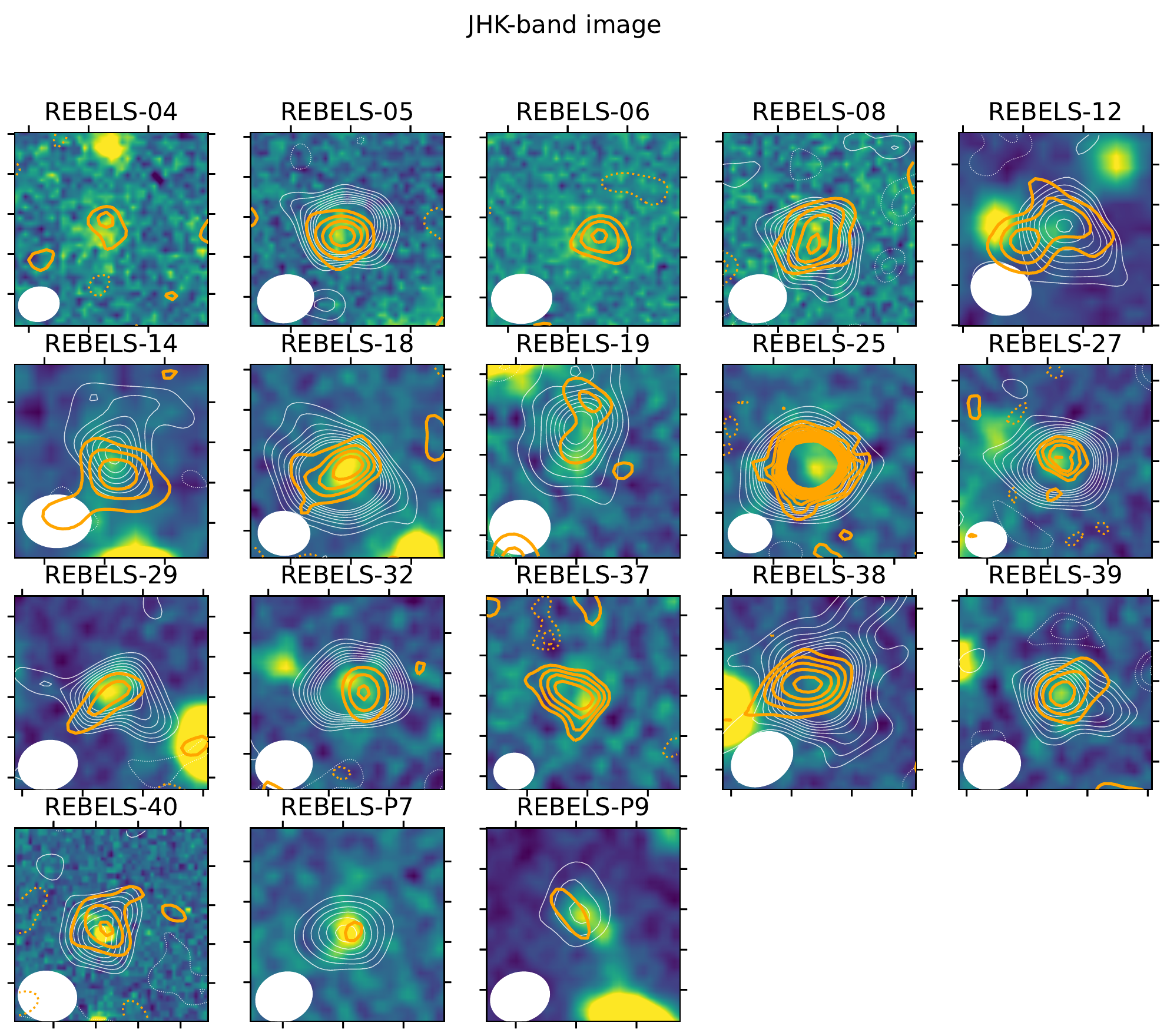}
  \caption{ Dust continuum emission (the orange contours) of the REBELS
    sources. The background images ($5\arcsec \times 5\arcsec$) are
    the stacked $JHK$-band images (or the stacked $JH$ images when
    $K$-band is not available). The \CII emission is also shown by the
    thin white contours (plotted only when there is a detection with 
    $\geq 3.3 \sigma$). The dotted contours indicate negative emission. 
    The contours start
    from $2\sigma$ in steps of $1\sigma$ of the background standard
    deviation.  The white ellipses at the bottom left are the beam
    size.
    \label{fig:dust_det}
  }
\end{figure*}

\section{Extraction of Dust Continuum Emission} \label{sec:meas}

\subsection{Identifying Dust Emission} \label{subsec:src_extr}

We used \texttt{PyBDSF} \citep{Moha15} to detect dust emission in each
ALMA dust continuum image without any primary beam correction applied.
\texttt{PyBDSF} identifies ``islands'' of adjacent emission and then
these ``islands'' are decomposed by fitting multiple Gaussian
functions to find individual sources and their peaks.  In this
process, \texttt{PyBDSF} uses two signal-to-noise (SNR) thresholds,
iSNR and pSNR, to determine the boundary of an island and the peak of
a source, respectively. These thresholds scale with the background
RMS, which we let the software calculate using the input image.
The minimum allowed island size was set to one-third of the number 
of pixels in the beam area (the default setting of \texttt{PyBDSF}).

The parameters iSNR and pSNR are crucial in ensuring robust
detections. In order to test the reliability of detected sources, we
used a range of combinations of iSNR and pSNR to perform source
extractions on the dust continuum images. The same test was also
performed on the reversed science images (hereafter negative images;
$-1 \times$\,science image). We employed the ranges of $1.0 \leq {\rm
  iSNR} \leq 3.0$ and $2.0 \leq {\rm pSNR} \leq 5.0$, divided into
steps of 0.1 for both of the parameters, but excluded the cases with
${\rm iSNR} > {\rm pSNR}$.  For each extraction, we counted the number
of detected sources in the science image as a number of positive
detections ($N_{\rm pos}$) and a number of negative detections
($N_{\rm neg}$) in the reversed image. These numbers were used to
evaluate the ``purity'' ($p$) defined as
\[ p = \frac{N_{\rm pos} - N_{\rm neg}}{N_{\rm pos}} \]
to indicate the reliability of the detections with a certain set of
iSNR and pSNR.

In Figure~\ref{fig:purity_2D}, we show 2D histograms (colour maps) of
$p$, $N_{\rm pos}$, and $N_{\rm neg}$ as functions of iSNR and pSNR
after combining all of the \texttt{PyBDSF} detections of the entire
REBELS sample.  Our primary interest is to detect dust emission of the
REBELS sources at the phase centre.  Thus, we limit the search area
for the positive sources in a circular area of $1.5\arcsec$ radius
centred in the pointing field.
The detections of negative sources, on the other hand, were performed
in the full primary beam area, in order to improve statistics of the
detections. Then, $N_{\rm neg}$ was scaled down to match the search
area of the positive sources.

The purity increases with both pSNR and iSNR. To maximise the number
of reliable detections, we adopted the detection thresholds of (pSNR,
iSNR)\,$= (3.3, 2.0)$, where the purity reaches
  $95\%$, for identifying the dust continuum of the REBELS targets.
We show $N_{\rm pos}$, $N_{\rm neg}$, and $p$ against pSNR (with
iSNR\,$=2.0$) in Figure~\ref{fig:purity_1D}.
The purity of the detections in the entire pointing field (including serendipitous
detections) is also displayed for comparison. In this case, $p=95\%$
is reached at pSNR\,$=4.2$.

In total we detected dust continuum emission with
  $\geq 3.3\sigma$ in \NDetCont out of the 40 REBELS targets ($\geq
40\%$~\footnote{This is currently a lower limit because out of the six
targets whose observations have not yet been completed, only three at
present have a dust continuum detection.})  in the central
$r=1.5\arcsec$ of the images. The detected sources are listed in 
Table~\ref{tbl:det_src} and displayed in
Figure~\ref{fig:dust_det} (We show the entire sample including 
non-detections in Figure~\ref{fig:dust_det_all}).
Among the \NDetCont dust continuum detected sources, three of them do
not yet have an emission line detection (REBELS-04, 06, and 37).
Further ALMA observations for these sources are underway.  The dust
continuum detections of non-primary target sources will be discussed
in a forthcoming paper.

\begin{center}
  \begin{table*}
    \caption{Compilation of current dust continuum
        detections of star-forming galaxies at $z > 6.5$ sorted by
        redshift. The photometric redshifts are indicated by 
        daggers. The two sources with asterisks are also 
        the REBELS targets, but we do not detect their dust continuum 
        emission (see \S\ref{sec:properties}). 
    \label{tbl:dust_z}}
    \begin{tabular}{lll}
      \hline
      Source name     &    Redshift    & References \\
      \hline  
      REBELS-04	      &     8.57$^\dag$ & This work \\
      A2744\_YD4      &     8.38        & \cite{Lapo17a} \\
      MACS0416\_Y1    &     8.31        & \cite{Tamu19,Bakx20} \\
      REBELS-37	      &     7.75$^\dag$ & This work \\
      REBELS-18	      &     7.67        & \cite{Scho21}; This work \\
      REBELS-19	      &     7.37        & This work \\
      REBELS-40	      &     7.36        & This work \\
      REBELS-12	      &     7.35        & This work \\
      REBELS-25	      &     7.31        & \cite{Scho21}; \Hygate; This work \\
      B14-65666       &     7.15        & \cite{Bowl18,Hash19a,Suga21} \\
      A1689-zD1       &     7.13        & \cite{Wats15,Knud17,Inou20,Bakx21} \\
      REBELS-27	      &     7.09        & \cite{Scho21}; This work \\
      REBELS-14	      &     7.08        & This work \\
      REBELS-P9	      &     7.06        & \cite{Bowl21b,Scho21}; This work \\
      ID238225$^{*}$ (REBELS-30)  &     6.98        & \cite{Bowl21b} \\
      REBELS-P8       &     6.85        & \cite{Scho21} \\ 
      REBELS-39	      &     6.84        & This work \\
      REBELS-06	      &     6.80$^\dag$ & This work \\
      REBELS-08	      &     6.75        & This work \\
      REBELS-P7	      &     6.75        & \cite{Scho21}; This work \\
      REBELS-32	      &     6.73        & This work \\
      REBELS-29	      &     6.68        & \cite{Bowl21b}; This work \\
      ID169850$^{*}$ (REBELS-34)  &     6.63        & \cite{Bowl21b} \\
      REBELS-38	      &     6.58        & This work \\
      REBELS-05	      &     6.50        & This work \\
      \hline  
    \end{tabular}
  \end{table*}
\end{center}

\begin{figure*}
  \begin{center}
    \includegraphics[width=0.49\textwidth]{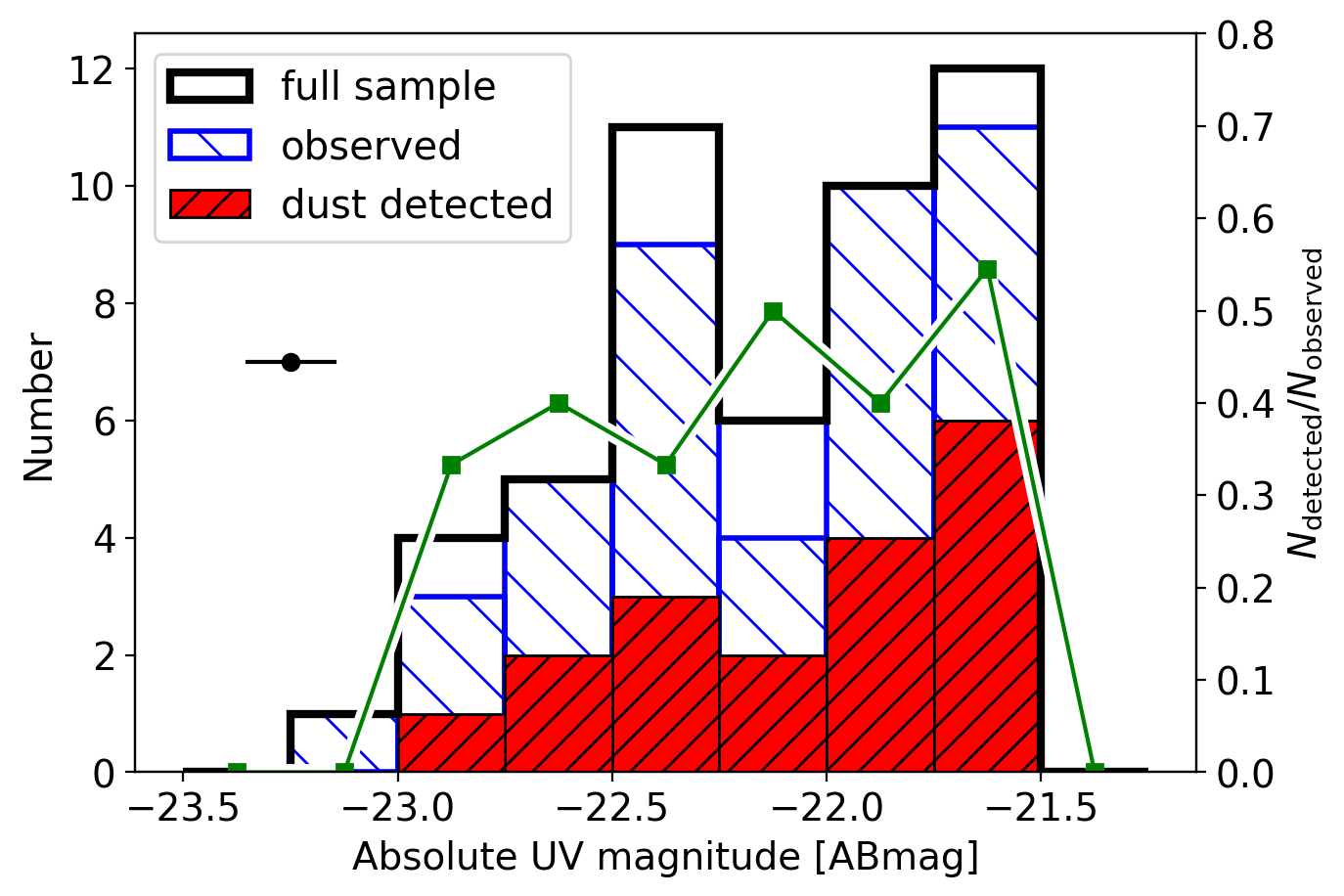}  
    \includegraphics[width=0.49\textwidth]{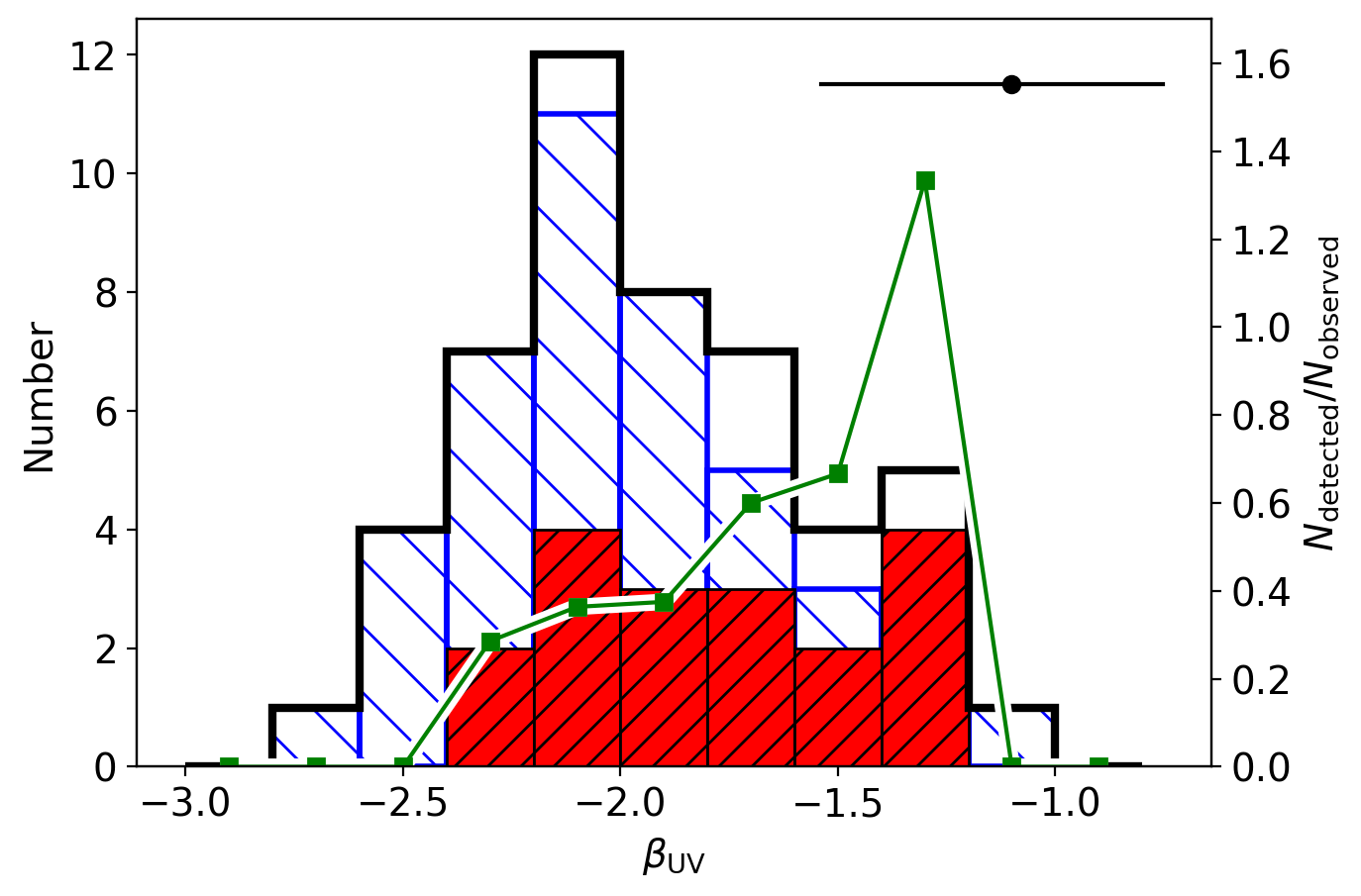}
    \\
    \includegraphics[width=0.49\textwidth]{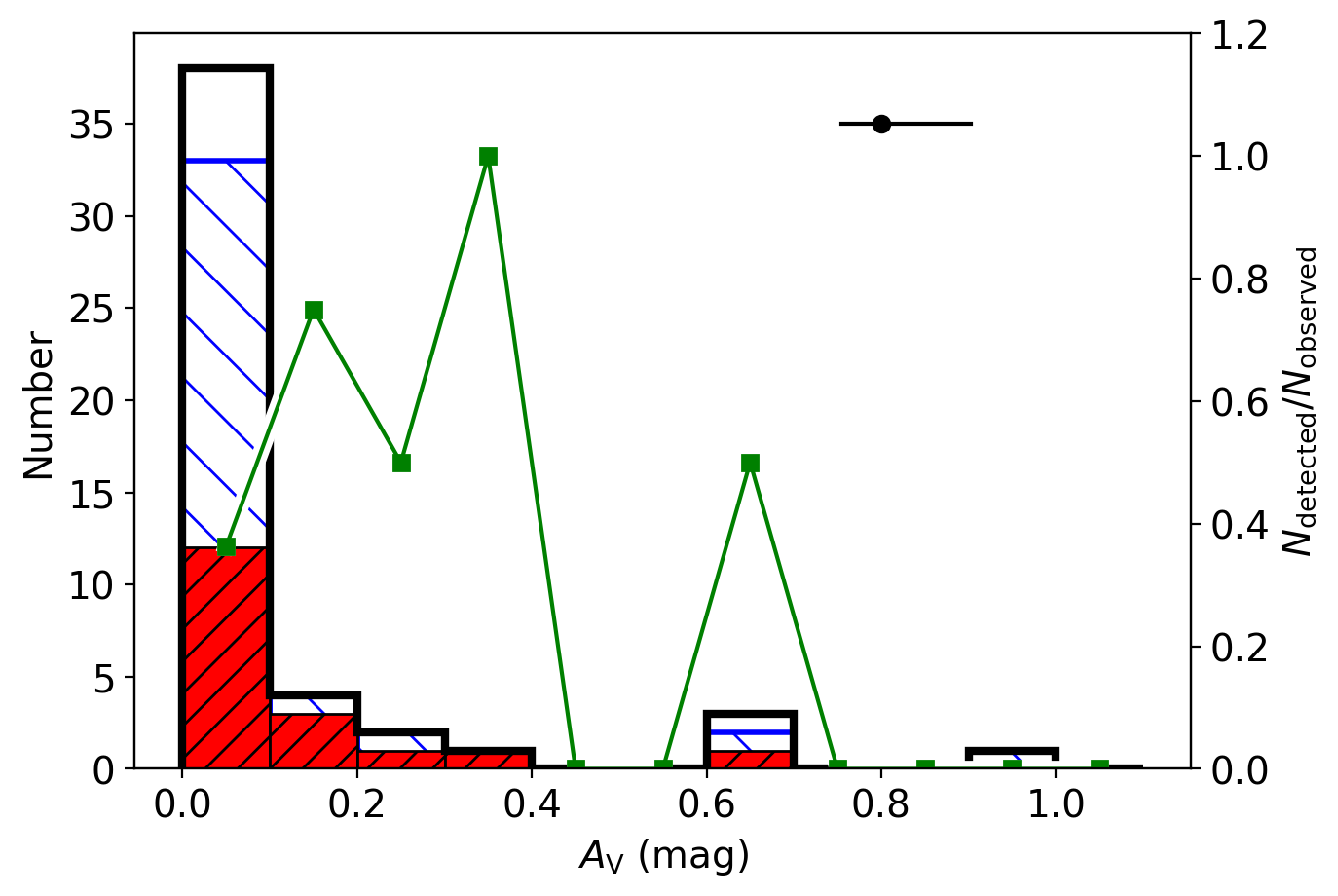}
    \includegraphics[width=0.49\textwidth]{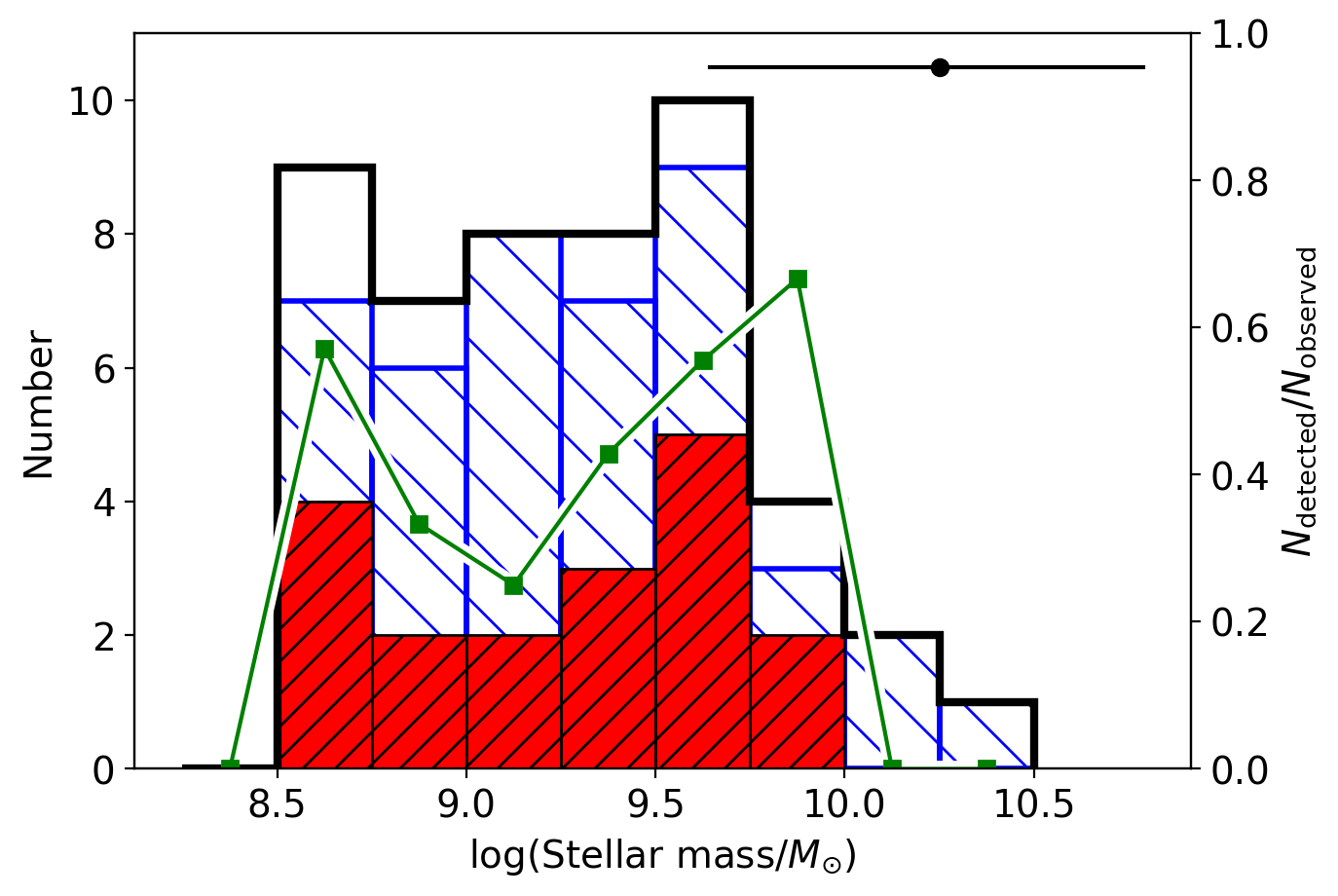}
    \caption{The physical properties ($M_{\rm UV}$, $\beta_{\rm UV}$,
      $A_V$, and $M_*$) as derived from SED fitting to the rest-UV 
      photometry of the REBELS sources.
      The typical error of the measured 
      parameters is shown by a black dot and a horizontal 
      line in each panel.
      The numbers of the REBELS sources with and without dust
      continuum detections are shown in the red and blue histograms,
      respectively. The black empty histogram denotes the entire
      REBELS sample, including the targets whose observations have not
      yet been completed. 
      The green line indicates
      the number fraction (the right axis) of the dust continuum 
      detected sources (red) 
      against the fully observed sources (red).
      \label{fig:dust_prop}   
    }
  \end{center}
\end{figure*}

\subsection{Dust Continuum Flux Measurements} \label{subsec:flux_meas}

Based on the dust emission identified with \texttt{PyBDSF}, we
measured the dust continuum flux densities at each location of the
detections.  We employed the \texttt{imfit} task in \texttt{CASA} to
fit one or more Gaussian components in the dust continuum images.
This is because there are some galaxies with complex
  morphology, for which \texttt{imfit} is preferred as an apriori
  specified number of Gaussians can be fitted.

We limited the fitting region to $3\arcsec \times
3\arcsec$~\footnote{Except for REBELS-40 (Super8-1), where a smaller
region ($1.5\arcsec \times 1.8\arcsec$) was used to avoid a noise peak
immediately to the east of the source.}
centred on the source position found by \texttt{PyBDSF} and let
\texttt{imfit} automatically specify initial parameter estimates of a
Gaussian function when there was only a single source.  The fitting
parameters were peak intensity, peak position, major and minor axis,
and position angle.  For REBELS-12 and REBELS-19, there are two sources 
detected by \texttt{PyBDSF}
within our detection radius ($1.5\arcsec$, see
\S\ref{subsec:src_extr}).  In these cases, we provided initial
estimates of the fitting parameters to fit the two sources
simultaneously. We have not yet found \CII emission or a counterpart
at shorter wavelengths for either of the secondary sources (the source
further from the phase centre with the optical counterpart).  It is
difficult to assess if these sources are actually at $z\sim7$ with
currently available datasets. In Table~\ref{tbl:det_src}, we report
the fluxes of each of the components and the total fluxes.  We
investigated the residual image of each source to confirm the fits
were adequate.

To determine whether the detected source is resolved or not, we
performed Monte Carlo simulations to account for the flux boosting
effect due to noise emission \citep[e.g.,][]{Alge21}.  In the science
images, we inserted in total 20,000 mock point (unresolved)
sources~\footnote{We performed 100 simulations for each science image,
with 5 inserted mock sources in each simulation.} and used the same
detection method, \texttt{PyBDSF}, to find the sources. The ratios of
recovered integrated and peak flux densities ($S_{\rm int}$ and
$S_{\rm peak}$, respectively) were compared against peak $S/N$ of the
sources. The excess from unity in the
recovered-to-inserted flux ratio is due to the noise
emission boosting the flux.
  At $S/N \sim 8$ ($S/N \sim
  4$), the probability that a source is robustly resolved with a 95\%
  confidence level requires a ratio of $S_{\rm int}/S_{\rm peak} >
  1.25$ ($S_{\rm int}/S_{\rm peak} \sim 2$), otherwise a ratio above
  unity is more likely due to underlying noise.  All of the $S_{\rm
  int}/S_{\rm peak}$ ratios of the REBELS sources indicate that they
are unresolved with a 95\% probability, except for
  REBELS-25 which is the brightest source in our
  sample. Note that both REBELS-12 and 19 are
  resolved into two components, but their individual components are
  not resolved.  Thus, in Table~\ref{tbl:det_src}, the reported flux
for REBELS-25 is the \texttt{imfit} integrated flux, but the
\texttt{imfit} peak fluxes for the rest of the sources.

For the REBELS targets without a dust continuum detection, we provide
an upper limit in Table~\ref{tbl:det_src}.  To derive the $3\sigma$
upper limit, we used the RMS noise in each dirty image and scaled it
up by a factor of three. Then we multiplied this by a flux boosting 
correction factor of 1.33 at $S/N=3$ from the same set of simulations 
described above, based on a comparison of the inserted and recovered 
peak flux densities. This allows us to avoid
underestimating a random noise peak which may appear
close to the expected location of a source.  This is
a reasonable estimate for our targets because it is likely most of the
REBELS sources are not resolved in our observations.

Finally, we considered the effect of the Cosmic Microwave Background
(CMB) following the prescription of \cite{daCu13b}.  As the CMB
temperature increases with redshift, it can affect the dust
temperature of high redshift galaxies with two competing effects:
boosting dust continuum emission and reducing background contrast
against the dust continuum.  We assumed a dust SED with a dust
emissivity spectral index ($\beta$) of
2.0~\footnote{We assumed Milky Way-like dust here,
  which is consistent with recent measurements of high redshift
  galaxies (\citealt{daCu21}, \citealt{Bowl18}, \citealt{Scho21},
  \citealt{Ferr22})} and a dust temperature (\Td) of
46\,K~\footnote{If \Td is 10\,K lower (higher), the
  correction factor would be 6\% higher (3\% lower).}. 
  This is
  the median \Td of the REBELS sources with both dust and \CII
  detections, calculated based on the method proposed by
  \cite{Somm21}~\footnote{Although this model assumed a
    Salpeter IMF ($1-100\,{\rm M_{\odot}}$ ) and a metallicity range of
    $0.3-1 Z_\odot$, the resulting dust SEDs are not significantly 
    affected.} and adopted to the REBELS sources in
  \cite{Somm22}.  The CMB correction factors increase the fluxes 
  in the range of $\sim 4-18\%$
depending on the redshift of each galaxy. Because we currently do not
have a robust dust temperature measurement, the continuum flux
measurements in Table~\ref{tbl:det_src} have not been corrected for
the CMB effect. However, the reported \LIR were calculated based on
the continuum fluxes corrected for the CMB effect.

\begin{figure}
  \begin{center}
    \includegraphics[width=0.49\textwidth]{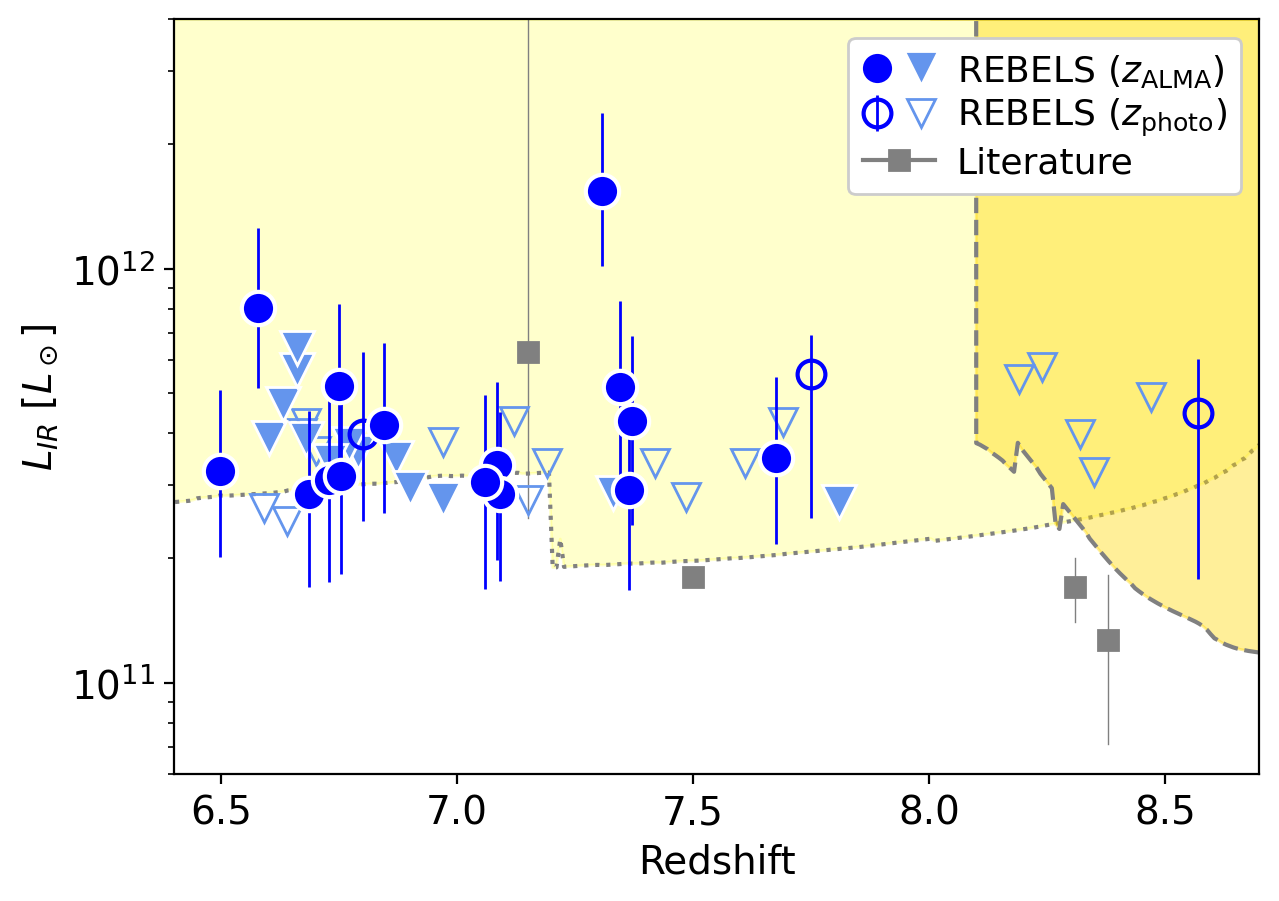}
    \caption{ \LIR vs. redshift for the REBELS sources and the earlier
      results from literature (Table~\ref{tbl:dust_z}).
      The dust continuum detected sources are displayed as the blue circles. 
      The light blue triangles are upper limits of the dust non-detected
      sources. For the sources with a \CII detection, their
      spectroscopic redshifts are used (the filled symbols), otherwise
      a photometric redshift is adopted (the open symbols).  
      The $3\sigma$ detection limits of the REBELS observations for 
      the continuum emission for the \CII
      and \OIII scans are shown in the background 
      as the light yellow region with the
      dotted line and the dark yellow region with the dashed line,
      respectively.
      \label{fig:LIR}
      }
  \end{center}
\end{figure}

\subsection{Total Infrared Luminosity} \label{subsec:LIR}

A \LIR conversion from a rest-frame 158\um (or
  88\um) continuum flux was used to acquire \LIR ($8-1000$\um) or its
  upper limit for all of the REBELS sources. The conversion factor
was computed from the infrared SEDs predicted by \cite{Somm22}.  This is
the same model that was used for deducing \Td above.

This model used the \CII line luminosity and its underlying dust
continuum emission to predict \Td. This model uses
  the \CII line luminosity as a proxy for the dust mass, given a
  [\ion{C}{ii}]-to-total gas conversion factor derived analytically
  assuming the Kennicutt–Schmidt relation \citep{Kenn98b} and the
  $L_{\rm [CII]}-$SFR relation \citep{DeLo14}.  Then, the underlying
  continuum emission at 158\um is used to constrain the \Td as well as
  the infrared SED of a single temperature modified blackbody. We
assumed the same $\beta=2.0$ as the one used for the CMB correction (see
\S\ref{subsec:flux_meas}).

For each REBELS source with both \CII and rest-frame 158\um continuum
detections, \LIR was computed using the SED derived with the method
above. Based on these \LIR and SEDs, a 158$\mu$m-based conversion
factor for each galaxy was then calculated. We adopted a single
conversion factor for all REBELS sources by taking a median
value. This allows us to provide \LIR (or its upper
  limit) for the galaxies without both \CII and dust continuum
  emission. The conversion factor is derived in \S3.3.1 in
\cite{Somm22} as:
\[ L_{\rm IR} = 14(^{+8}_{-5}) \, \nu L_{\nu, 158\mu{\rm m}}. \]
The obtained \LIR (or the \LIR upper limits) are reported in
Table~\ref{tbl:det_src}.

For a simple comparison, if we instead adopted the
  SED templates constructed with star-forming galaxies at lower
  redshifts \citep[$1 \lesssim z \lesssim 3$; e.g.,][]{Beth17,
    Schr18a}, we would obtain a slightly lower conversion factor
  ($\sim 11$) for $L_{\nu, 158\mu{\rm m}}$.  Furthermore, if we used a
  modified black body SED with $\beta=1.6$ and \Td$=54$\,K, which
  assumed a dust temperature evolution extrapolated to $z=7$
  \citep{Bouw20}, then this conversion factor, $11 \pm 3$, would also be
  consistent within the uncertainty of the conversion factor
  used in the current work.

There are four REBELS sources for which we targeted \OIII instead of
\CII, and thus the rest-frame continuum detection is at rest-frame
88\um.  For these cases, the conversion factor, 
  $L_{\rm IR} = 8(^{+1}_{-4}) \, \nu L_{\nu, 88\mu{\rm m}}$, was
derived from the same SED with the median \Td of the
the sample that has both the rest-frame 158\um continuum and 
\CII detections. Note that this conversion factor yields
about a factor of three higher $L_{\rm IR}$ than when a simple modified
blackbody (with $T_{\rm d}=46$K and $\beta=2.0$) is used to 
fit the single data point at rest-frame 88\um. This suggests that
the dust properties of the \CII detected galaxies at $z \sim 7$ and 
the 88\um targets at $z \gtrsim 8$ may be different. In fact, 
the two previously and only known dust continuum detected galaxies at 
$z > 8$ \citep[][see also Table~\ref{tbl:dust_z}]{Lapo17a,Bakx20} show 
different SEDs from the median SED adopted in this work. 
Both of their dust temperatures are estimated to be higher than 
$\sim 90$\,K. If this is also the case for our $z \gtrsim 8$ sample, 
then these galaxies would have a higher $L_{\rm IR}$ and 
a lower dust mass than the currently assumed SED that is used to
derive the conversion factor.

\section{Galaxies with Dust Emission at $\mathbf{z \sim 7}$} \label{sec:properties}

Although the REBELS data are still not fully acquired,
we have detected significant dust continuum emission in \NDetCont
REBELS targets and two from the pilot programs \citep[known as
  UVISTA-Z-019 and UVISTA-Z-001, respectively, in][]{Scho21}.  These
two detections are included in the analyses of this paper.
There are four galaxies that overlap with \cite{Bowl21b}. They reported
dust continuum detections in Band 6 for REBELS-29 and REBELS-34, a marginal
detection ($\sim 3\sigma$) for REBELS-30, and no detection for
REBELS-31, whereas we do not find a detection in REBELS-34 and REBELS-30 with
the REBELS data. This is likely due to a shallower depth of the REBELS
observations compared to the program of \cite{Bowl21b}. 
Our upper limit estimates are consistent with their flux measurements.  In
Table~\ref{tbl:det_src}, we include their flux measurements for
completeness. These additional detections lead to a total of 20 dust
continuum detections out of the 49 targets of the REBELS and the pilot
program ($\geq 40\%$). This is still a lower limit to
  what we expect from the full REBELS dataset once completed.  Among
the dust continuum detected galaxies, 15 have been spectroscopically
confirmed with the \CII line at $z > 6.5$.

Together with the pilot observations \citep{Scho21}, REBELS has
increased the number of dust continuum detections of star-forming
galaxies at $z\sim7$ by at least a factor of three compared with the
previously known detections (See Table~\ref{tbl:dust_z}).
There are still six sources whose observations have either been
partially executed or not yet been executed.  Among the targets with
partial observations, the dust continuum emission is already detected
for REBELS-04, 06, and 37, but without any line emission. Based on
their photometric redshifts, we are aiming to detect \CII for 
REBELS-06 ($z_{\rm photo} = 6.80^{+0.13}_{-0.11}$) and \OIII for 
REBELS-04 ($z_{\rm photo} = 8.57^{+0.10}_{-0.09}$) and 
REBELS-37 ($z_{\rm photo} = 7.75^{+0.09}_{-0.17}$).  
The detection rate of at least 40\% in the REBELS sample suggests 
that it is not uncommon for galaxies at $z \sim 7$ to have 
significant dust continuum emission and host obscured star formation.

In the following subsections, we discuss the rest-frame UV and
infrared properties of the dust continuum detected galaxies in the
REBELS sample. We adopted a prescription of 
${\rm SFR_{UV}} = 7.1\times10^{-29}L_\nu$ [$\rm M_\odot\,yr^{-1}\,(erg\,s^{-1}\,Hz^{-1})^{-1}$] and 
${\rm SFR_{IR}} = 1.2 \times 10^{-10}L_{\rm IR}$ 
[$\rm M_\odot\,yr^{-1}\,L_\odot^{-1}$] \citep[the same as ][]{Bouw21b}.
The derived SFRs are summarised in Table~\ref{tbl:SFR}.

\begin{table}
\centering
\caption{SFRs derived based on UV and IR, and the total (UV+IR) for the REBELS sample.
\label{tbl:SFR}}
\begin{tabular}{cccc}
  \hline
  REBELS ID &         SFR$_{\rm UV}$ &         SFR$_{\rm IR}$ &      SFR$_{\rm UV+IR}$ \\ 
            & $\rm M_\odot\,yr^{-1}$ & $\rm M_\odot\,yr^{-1}$ & $\rm M_\odot\,yr^{-1}$ \\ 
   \hline
  REBELS-01 & $    43\,^{+ 5}_{- 4}$ & $<                 34$ & $<                 78$ \\ 
  REBELS-02 & $    22\,^{+ 4}_{- 4}$ & $<                 29$ & $<                 51$ \\ 
  REBELS-03 & $    16\,^{+ 4}_{- 4}$ & $<                 33$ & $<                 50$ \\ 
  REBELS-04 & $    26\,^{+ 1}_{- 1}$ & $  54\,^{+ 19}_{- 32}$ & $  80\,^{+ 19}_{- 32}$ \\ 
  REBELS-05 & $    13\,^{+ 3}_{- 2}$ & $  39\,^{+ 22}_{- 15}$ & $  52\,^{+ 22}_{- 15}$ \\ 
  REBELS-06 & $    15\,^{+ 4}_{- 3}$ & $  48\,^{+ 28}_{- 18}$ & $  63\,^{+ 28}_{- 19}$ \\ 
  REBELS-07 & $    21\,^{+ 6}_{- 5}$ & $<                 33$ & $<                 54$ \\ 
  REBELS-08 & $    16\,^{+ 7}_{- 5}$ & $  63\,^{+ 36}_{- 24}$ & $  79\,^{+ 37}_{- 25}$ \\ 
  REBELS-09 & $    50\,^{+16}_{-12}$ & $<                 41$ & $<                 91$ \\ 
  REBELS-10 & $    37\,^{+12}_{- 9}$ & $<                 41$ & $<                 78$ \\ 
  REBELS-11 & $    39\,^{+ 9}_{- 7}$ & $<                 69$ & $<                109$ \\ 
  REBELS-12 & $    30\,^{+10}_{- 7}$ & $  62\,^{+ 38}_{- 27}$ & $  92\,^{+ 39}_{- 28}$ \\ 
  REBELS-13 & $    44\,^{+10}_{- 8}$ & $<                 65$ & $<                109$ \\ 
  REBELS-14 & $    36\,^{+16}_{-11}$ & $  40\,^{+ 24}_{- 17}$ & $  76\,^{+ 28}_{- 20}$ \\ 
  REBELS-15 & $    33\,^{+11}_{- 8}$ & $<                 43$ & $<                 76$ \\ 
  REBELS-16 & $    30\,^{+ 2}_{- 2}$ & $<                 44$ & $<                 74$ \\ 
  REBELS-17 & $    14\,^{+ 3}_{- 3}$ & $<                 47$ & $<                 61$ \\ 
  REBELS-18 & $    27\,^{+ 4}_{- 3}$ & $  42\,^{+ 24}_{- 16}$ & $  69\,^{+ 24}_{- 16}$ \\ 
  REBELS-19 & $    14\,^{+ 3}_{- 3}$ & $  51\,^{+ 31}_{- 22}$ & $  65\,^{+ 31}_{- 23}$ \\ 
  REBELS-20 & $    17\,^{+ 2}_{- 2}$ & $<                 52$ & $<                 68$ \\ 
  REBELS-21 & $    18\,^{+ 4}_{- 3}$ & $<                 32$ & $<                 50$ \\ 
  REBELS-22 & $    25\,^{+ 3}_{- 2}$ & $<                 34$ & $<                 58$ \\ 
  REBELS-23 & $    14\,^{+ 9}_{- 5}$ & $<                 47$ & $<                 61$ \\ 
  REBELS-24 & $    20\,^{+ 5}_{- 4}$ & $<                 39$ & $<                 59$ \\ 
  REBELS-25 & $    14\,^{+ 3}_{- 3}$ & $ 185\,^{+101}_{- 63}$ & $ 199\,^{+101}_{- 63}$ \\ 
  REBELS-26 & $    17\,^{+ 2}_{- 2}$ & $<                 56$ & $<                 73$ \\ 
  REBELS-27 & $    18\,^{+ 4}_{- 4}$ & $  34\,^{+ 20}_{- 13}$ & $  52\,^{+ 20}_{- 14}$ \\ 
  REBELS-28 & $    30\,^{+ 9}_{- 7}$ & $<                 42$ & $<                 72$ \\ 
  REBELS-29 & $    24\,^{+ 3}_{- 3}$ & $  34\,^{+ 20}_{- 14}$ & $  59\,^{+ 20}_{- 14}$ \\ 
  REBELS-30 & $    27\,^{+ 2}_{- 2}$ & $<                 35$ & $<                 62$ \\ 
  REBELS-31 & $    26\,^{+ 4}_{- 4}$ & $<                 51$ & $<                 76$ \\ 
  REBELS-32 & $    14\,^{+ 2}_{- 2}$ & $  37\,^{+ 23}_{- 16}$ & $  51\,^{+ 23}_{- 16}$ \\ 
  REBELS-33 & $    13\,^{+ 2}_{- 2}$ & $<                 48$ & $<                 61$ \\ 
  REBELS-34 & $    30\,^{+ 2}_{- 2}$ & $<                 45$ & $<                 75$ \\ 
  REBELS-35 & $    31\,^{+ 3}_{- 3}$ & $<                 46$ & $<                 77$ \\ 
  REBELS-36 & $    23\,^{+ 5}_{- 4}$ & $<                 33$ & $<                 56$ \\ 
  REBELS-37 & $    24\,^{+ 1}_{- 1}$ & $  67\,^{+ 16}_{- 37}$ & $  91\,^{+ 16}_{- 37}$ \\ 
  REBELS-38 & $    17\,^{+ 4}_{- 4}$ & $  96\,^{+ 54}_{- 35}$ & $ 114\,^{+ 54}_{- 35}$ \\ 
  REBELS-39 & $    38\,^{+ 6}_{- 5}$ & $  50\,^{+ 29}_{- 20}$ & $  88\,^{+ 30}_{- 20}$ \\ 
  REBELS-40 & $    17\,^{+ 1}_{- 1}$ & $  35\,^{+ 21}_{- 15}$ & $  52\,^{+ 21}_{- 15}$ \\ 
  REBELS-P1 & $    25\,^{+ 3}_{- 2}$ & $<                 41$ & $<                 66$ \\ 
  REBELS-P2 & $    15\,^{+ 6}_{- 4}$ & $<                 59$ & $<                 74$ \\ 
  REBELS-P3 & $    14\,^{+ 5}_{- 4}$ & $<                 51$ & $<                 65$ \\ 
  REBELS-P4 & $    18\,^{+ 9}_{- 6}$ & $<                 48$ & $<                 66$ \\ 
  REBELS-P5 & $    25\,^{+ 3}_{- 3}$ & $<                 41$ & $<                 65$ \\ 
  REBELS-P6 & $    16\,^{+ 2}_{- 2}$ & $<                 68$ & $<                 84$ \\ 
  REBELS-P7 & $    14\,^{+ 2}_{- 2}$ & $  38\,^{+ 23}_{- 16}$ & $  52\,^{+ 23}_{- 16}$ \\ 
  REBELS-P8 & $    17\,^{+ 2}_{- 1}$ & $<                 77$ & $<                 94$ \\ 
  REBELS-P9 & $    49\,^{+ 3}_{- 3}$ & $  37\,^{+ 23}_{- 16}$ & $  86\,^{+ 23}_{- 17}$ \\ 
  \hline
\end{tabular}
\end{table}

\subsection{Physical properties of the dust-detected galaxies at
  $\mathbf{z \sim 7}$}\label{subsec:UVprop}

Armed with our large sample of $z > 6.5$ galaxies with multiple
rest-UV and optical detections, 
we show the physical properties of the galaxies with and 
without dust continuum detections in Figure~\ref{fig:dust_prop}.

The number fractions of the dust continuum detected
sources are $\sim 0.4$ across the $M_{\rm UV}$ range with a hint of 
an increasing trend towards the less UV luminous end.
Although uncertainties are large, the UV continuum slope ($\beta_{\rm UV}$, 
defined as $L_\lambda \propto \lambda^\beta_{\rm UV}$) of the dust-detected 
sources tends to be redder (larger $\beta_{\rm UV}$) relative to the
non-detections. Almost all of the observed sources that are detected
with dust continuum have $\beta_{\rm UV} > -1.8$, but
there are two exceptions, REBELS-10 and 13. They have a red UV slope 
($\beta_{\rm UV} > -1.5$) but no dust detection. 
Neither of them have a \CII detection. Additional analyses reveal 
that these two sources may well be lower redshift interlopers. 
Newly available deeper {\it Spitzer}/IRAC observations \citep{Stef19sptz.prop}, 
which became available after the targeting of REBELS-10, put the photometric 
redshift of this source at $z < 6$. 
For REBELS-13, the integrated redshift likelihood distribution at $z < 6$ 
is $>15\%$. Detailed discussions on each of these sources will be presented 
in \Stefanon. More detailed analyses of $\beta_{\rm UV}$ and dust
properties (e.g., the IRX-$\beta_{\rm UV}$ relation) will be presented
in Bowler et al. (in prep).

On the other hand, we do not see a clear trend in the dust extinction
in the V-band ($A_V$) and stellar mass ($M_*$) from the SED
fitting. The limited rest-frame optical/NIR coverage we have of targets 
results in large uncertainties on estimates of the stellar mass 
\citep[see also][]{Topp22}. None of these relations are
statistically significant ($p$-value $> 0.05$) with the two-sample 
Kolmogorov–Smirnov test to reject the null-hypothesis that the distributions
of the dust continuum detected and undetected samples are identical.
Future observations with the {\it James Webb} Space Telescope ({\it JWST}) 
are essential for better constraints on these fundamental properties of 
$z\sim7$ galaxies (cf. the Cycle-1 General Observer Program ID 1626, 
PI: Stefanon).

\begin{figure*}
  \begin{center}
    \includegraphics[width=0.49\textwidth]{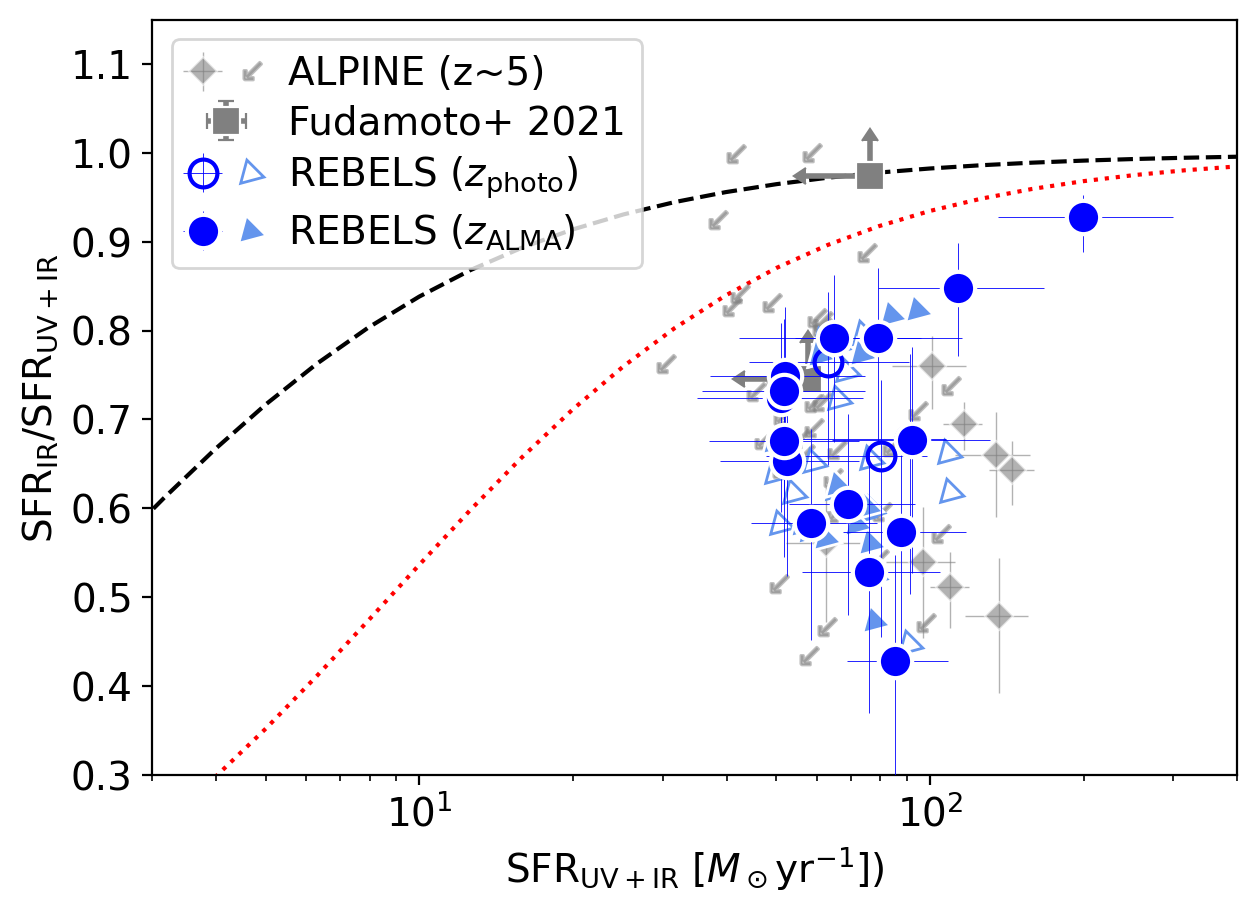}
    \includegraphics[width=0.49\textwidth]{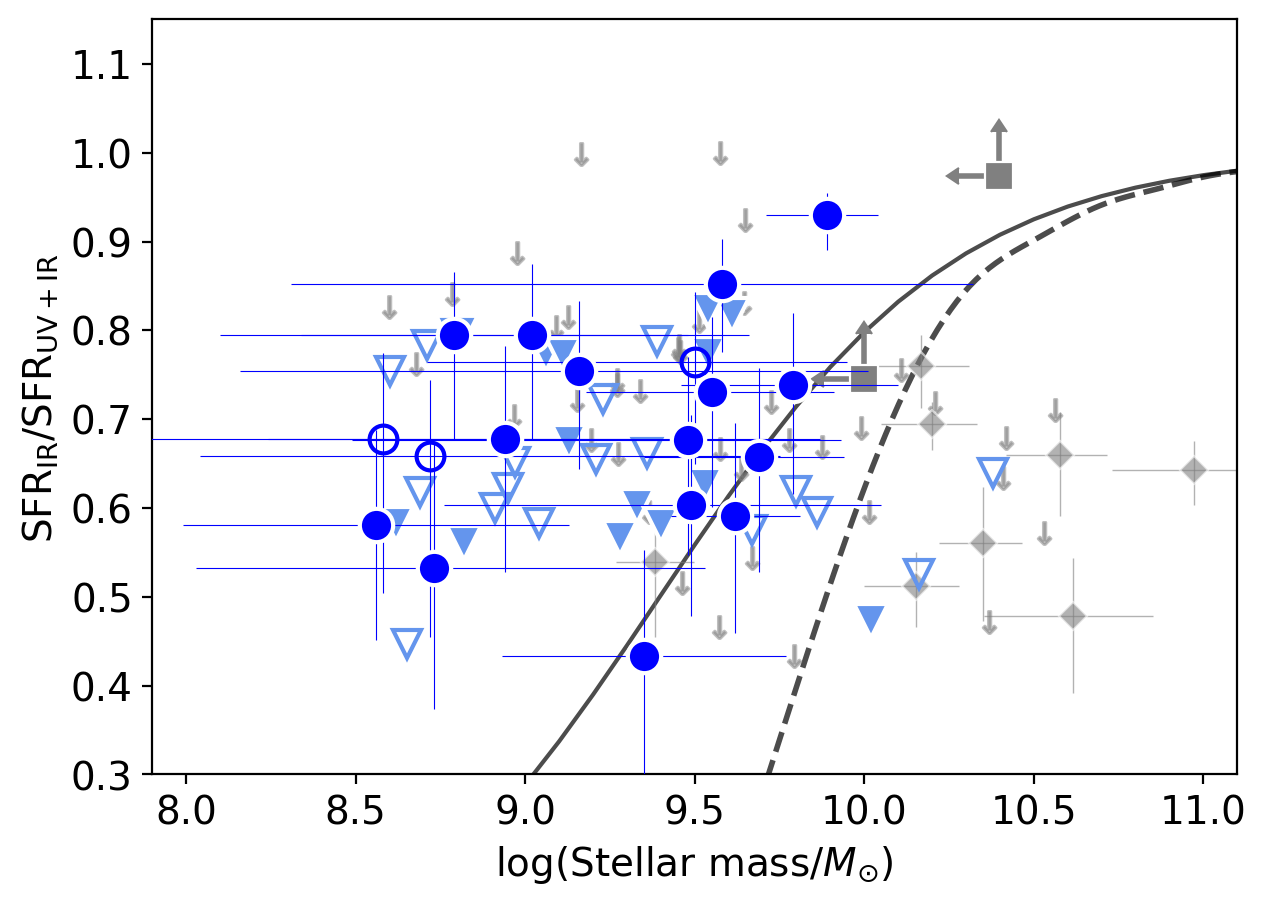}
    \caption{Fraction of obscured star formation rate
      (SFR$_{\rm IR}$/SFR$_{\rm UV+IR}$) as a function of total SFR (left) and
      stellar mass (right). The symbols are the same as in
      Figure~\ref{fig:LIR}. The limits of the two
        serendipitously detected $z\sim7$ galaxies are shown by the
        filled grey squares \citep{Fuda21}. 
        The $z\sim5$ ALPINE sources \citep{Fuda20b} are shown by
        diamonds for the continuum detected and by arrows as 
        the upper limits for the non-detected sources.
        In the left panel, the
      black dashed and red dotted lines indicate the relation at 
      $0.5 < z < 1.0$ and $2.0 < z < 2.5$, respectively 
      \citep[][with the \protect\cite{Dale02}
        SED templates]{Whit17}. In the right panel, the black solid
      line and dashed line represent the relations with the SED
      templates of \protect\cite{Dale02} and 
      \protect\cite{Kirk15}/\protect\cite{Magd12b},
      respectively, at $0.5 < z < 2.5$ where no evolution is seen
      \citep{Whit17}. 
      \label{fig:fobs}
      }
  \end{center}
\end{figure*}

\begin{figure*}
  \begin{center}
    \includegraphics[scale=0.26, clip, trim=20 10 60 20]{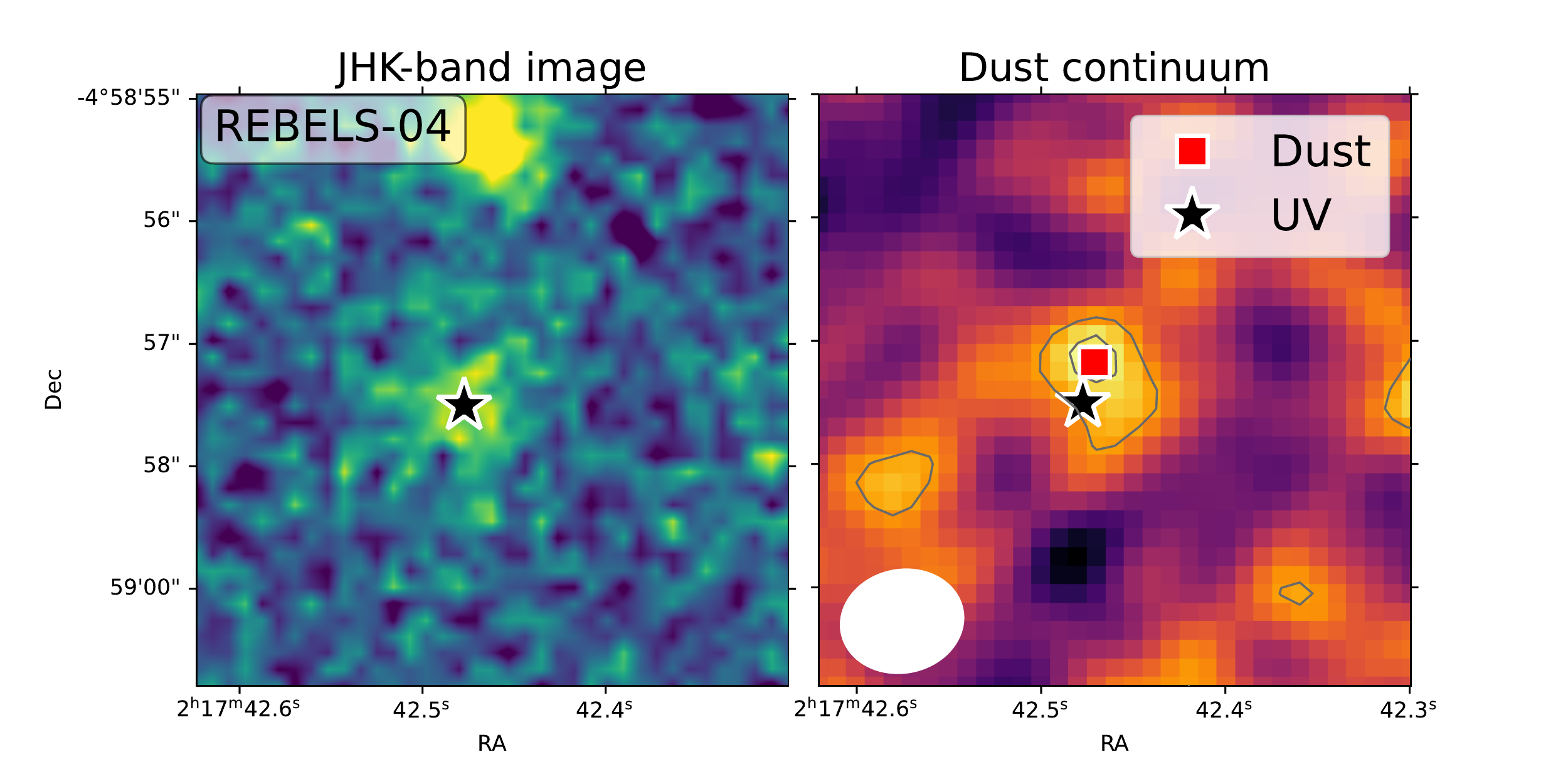}
    \includegraphics[scale=0.26, clip, trim=20 10 60 20]{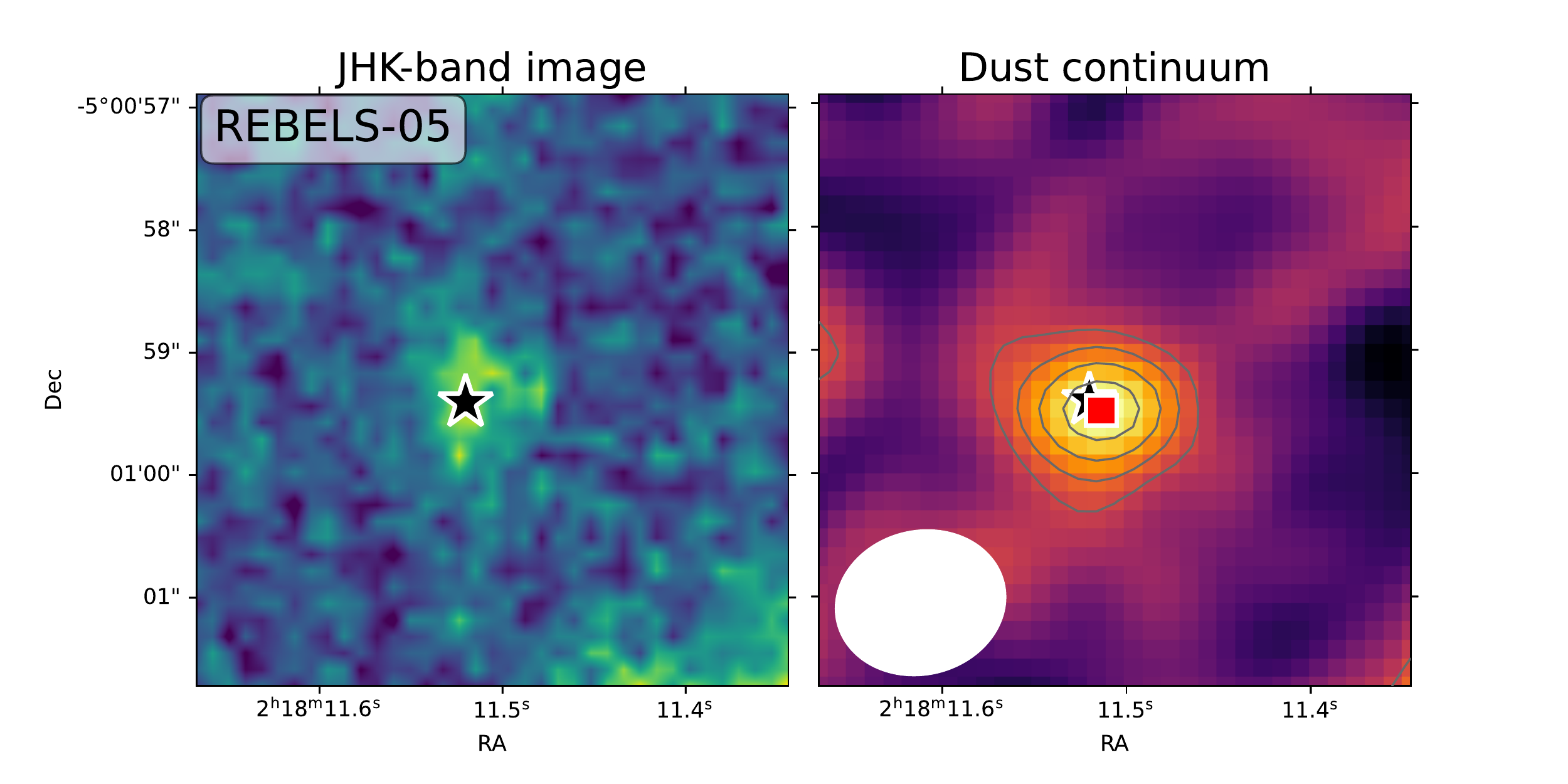}
    \includegraphics[scale=0.26, clip, trim=20 10 60 20]{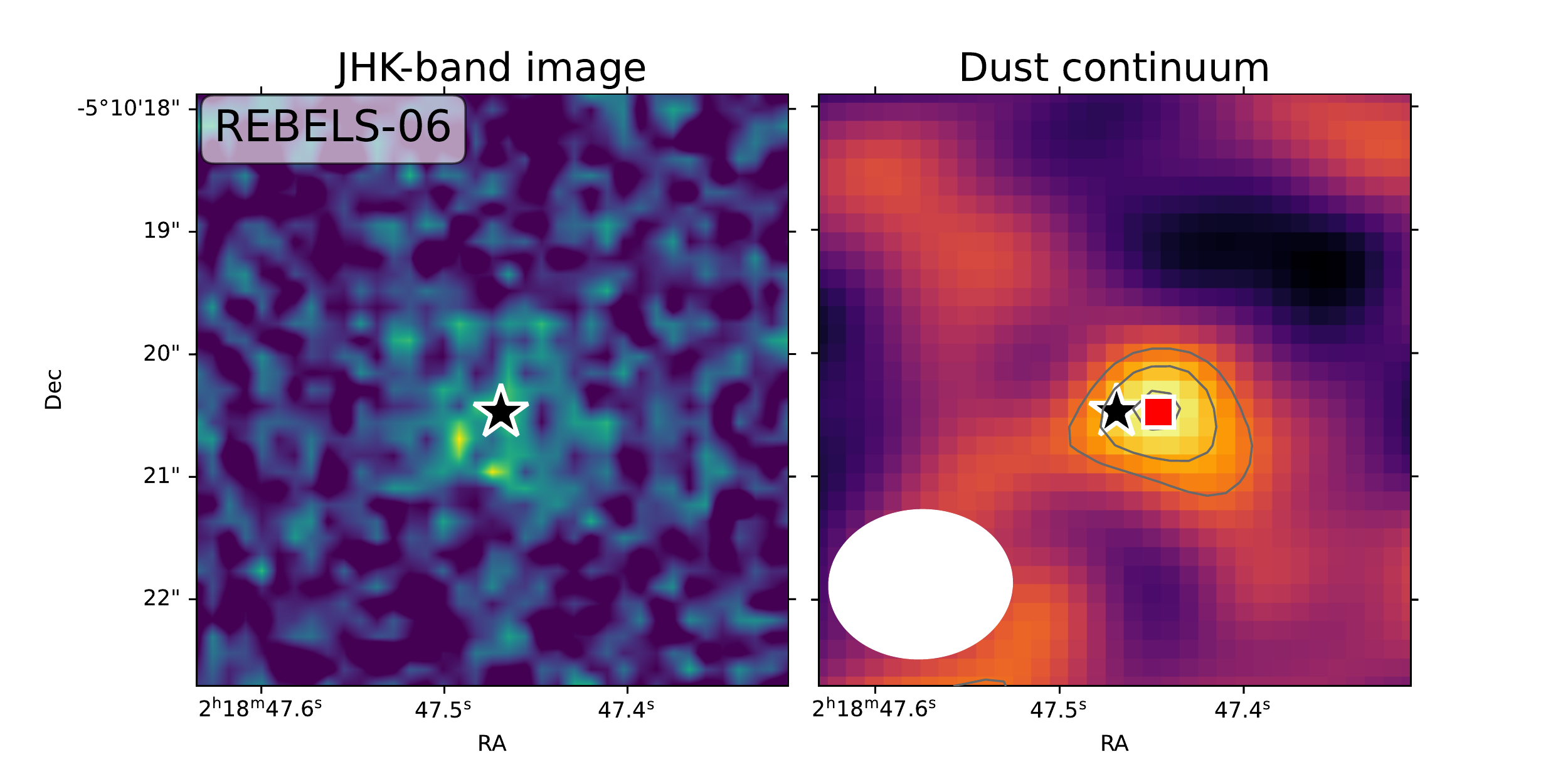}
    \includegraphics[scale=0.26, clip, trim=20 10 60 20]{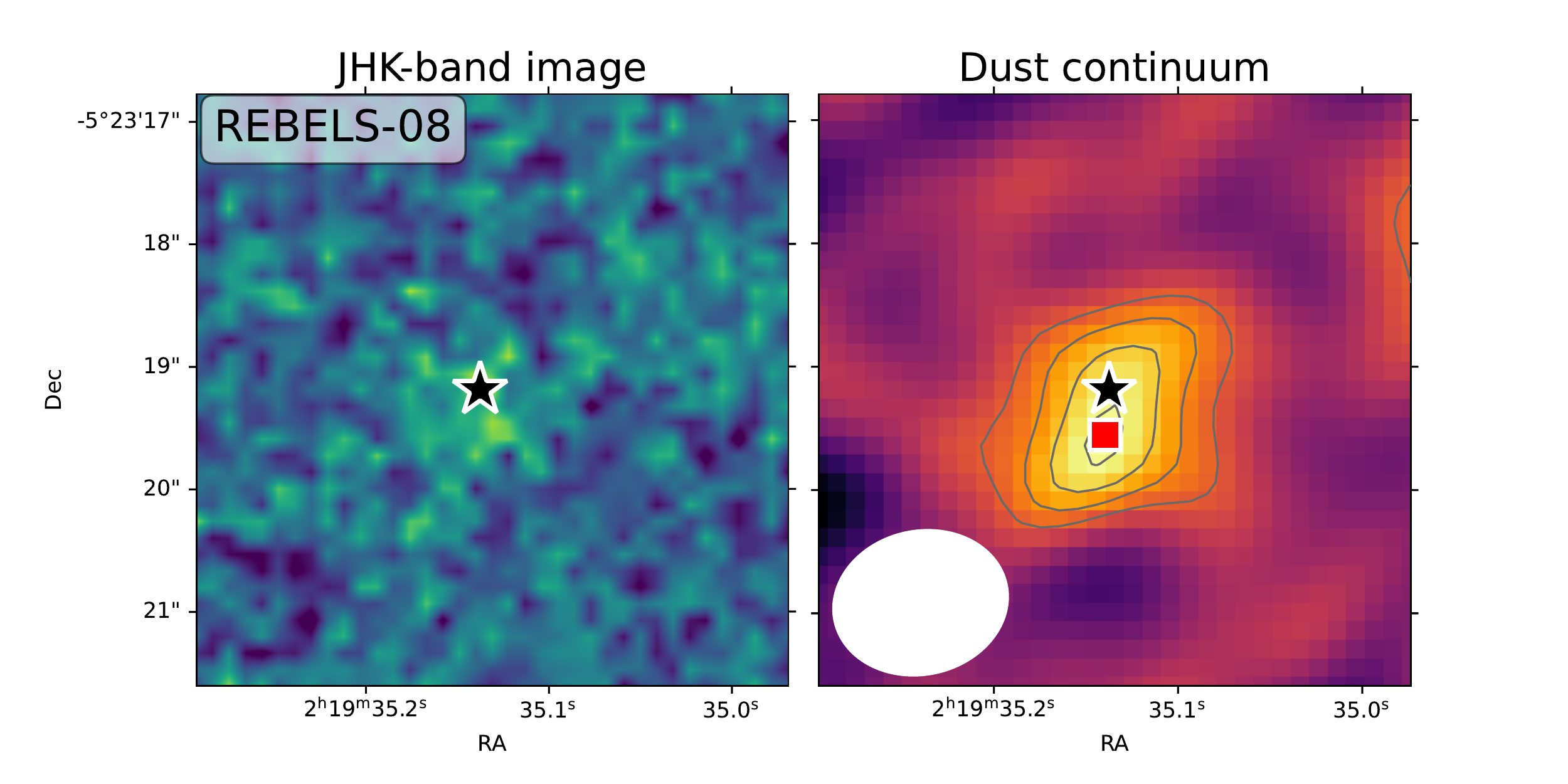}
    \includegraphics[scale=0.26, clip, trim=20 10 60 20]{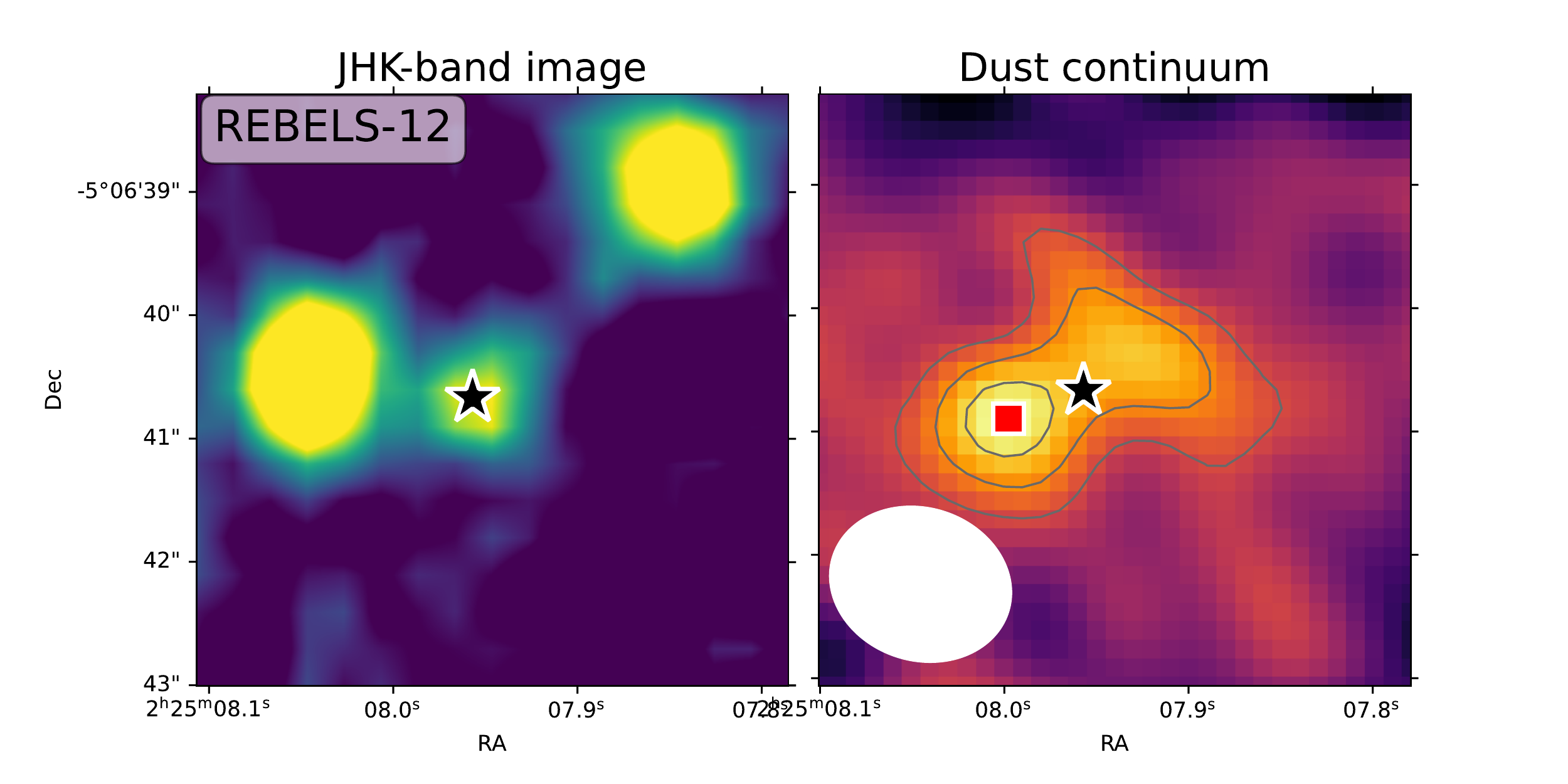}
    \includegraphics[scale=0.26, clip, trim=20 10 60 20]{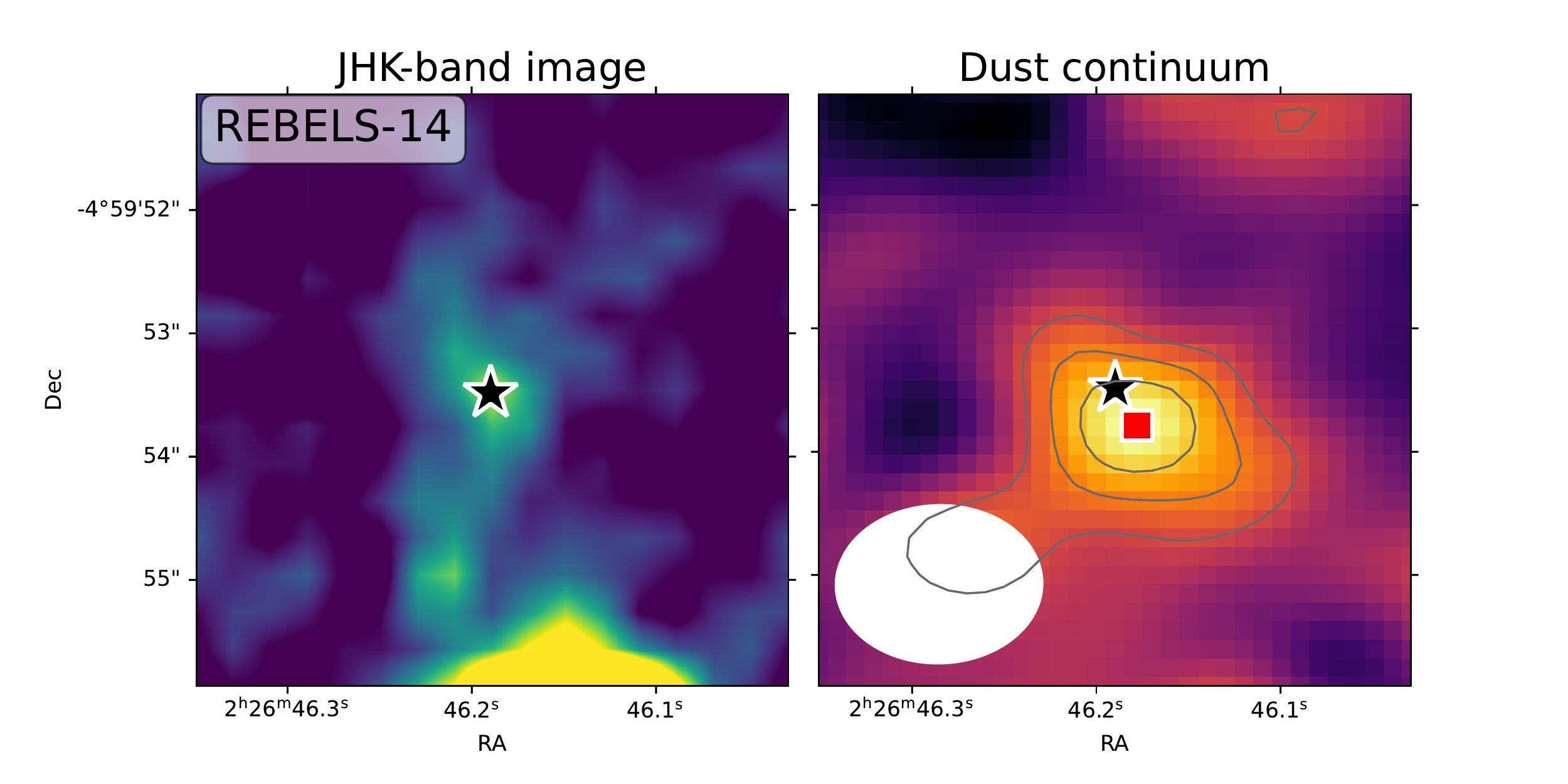}
    \includegraphics[scale=0.26, clip, trim=20 10 60 20]{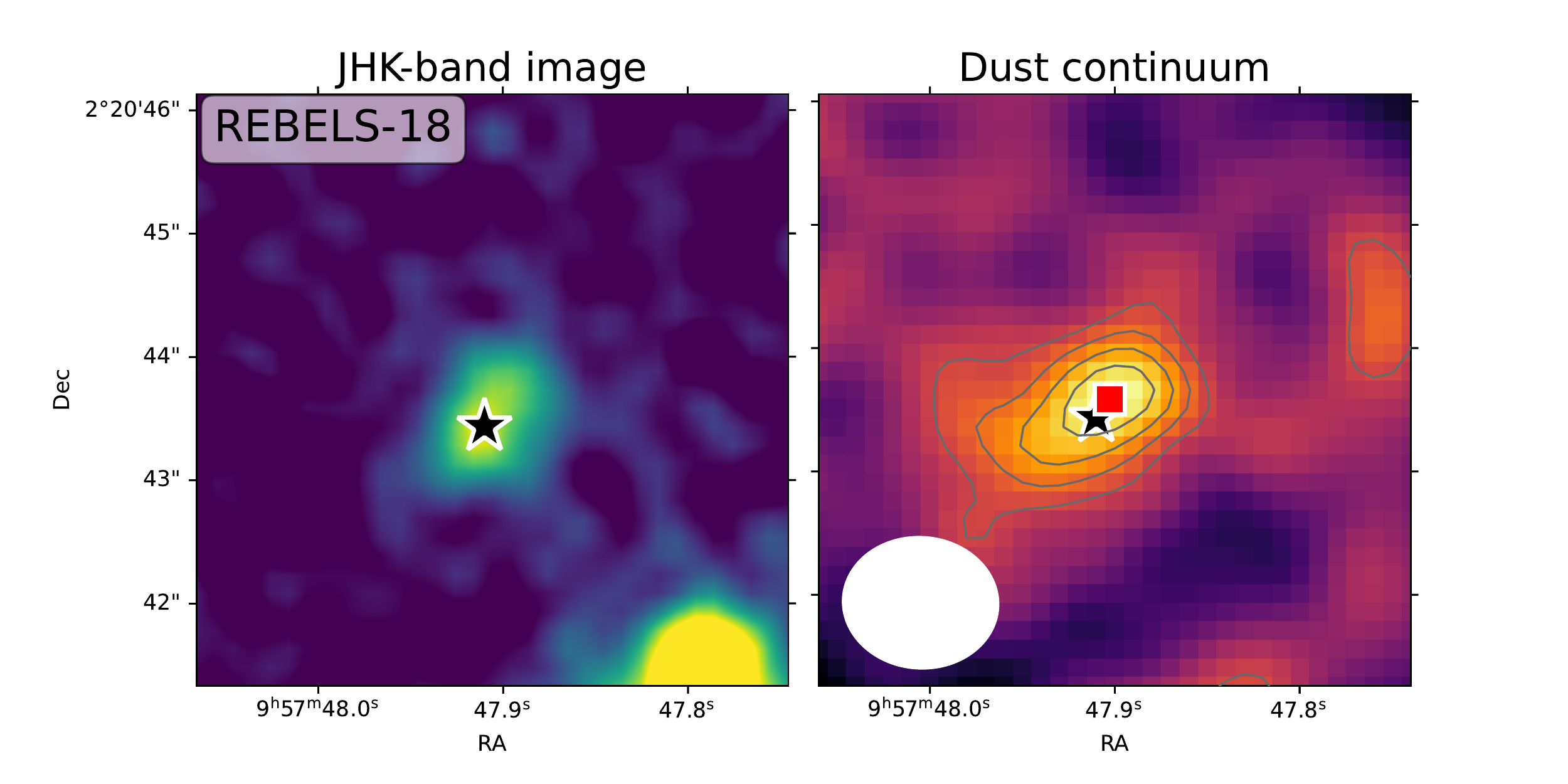}
    \includegraphics[scale=0.26, clip, trim=20 10 60 20]{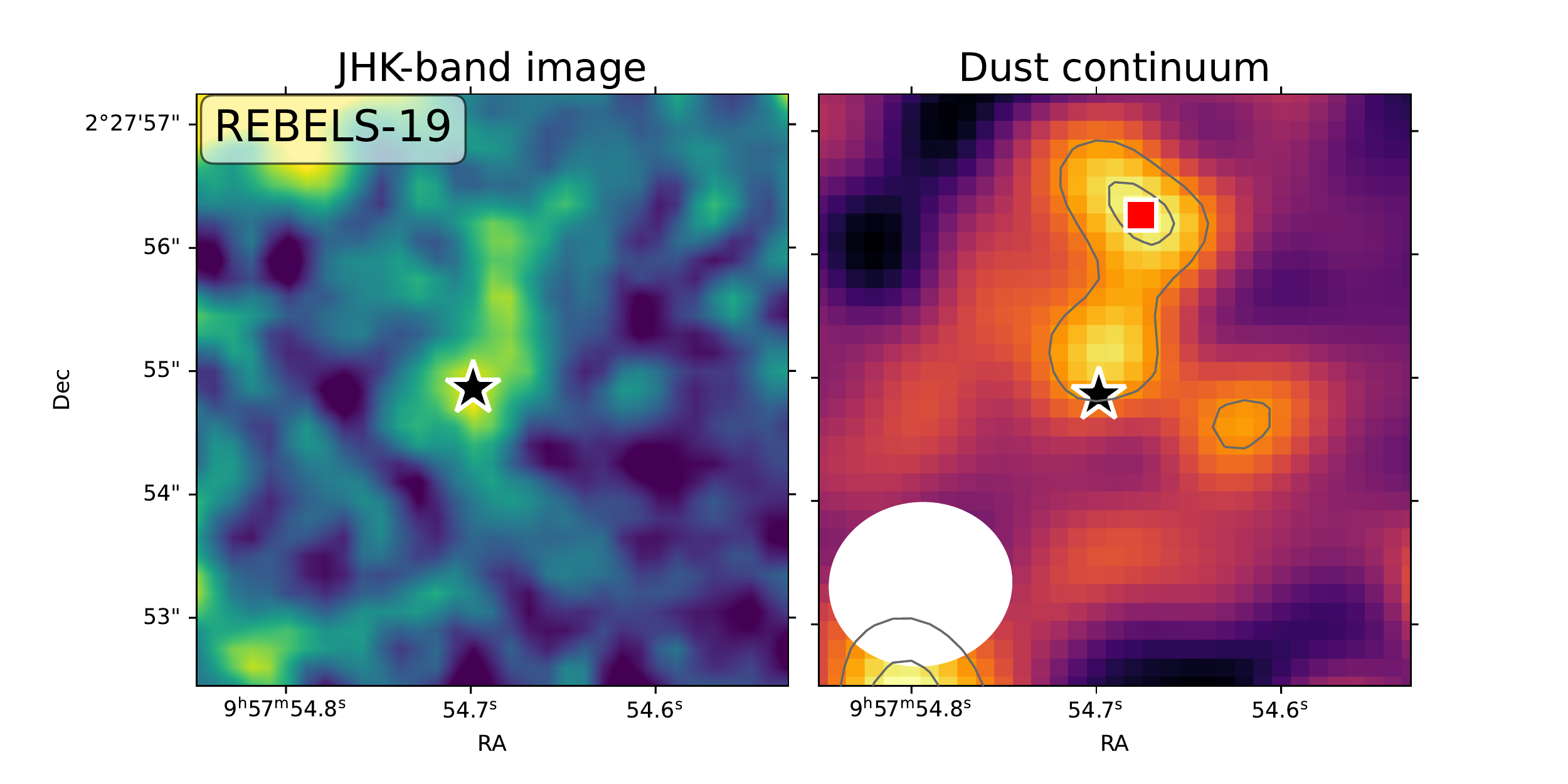}
    \includegraphics[scale=0.26, clip, trim=20 10 60 20]{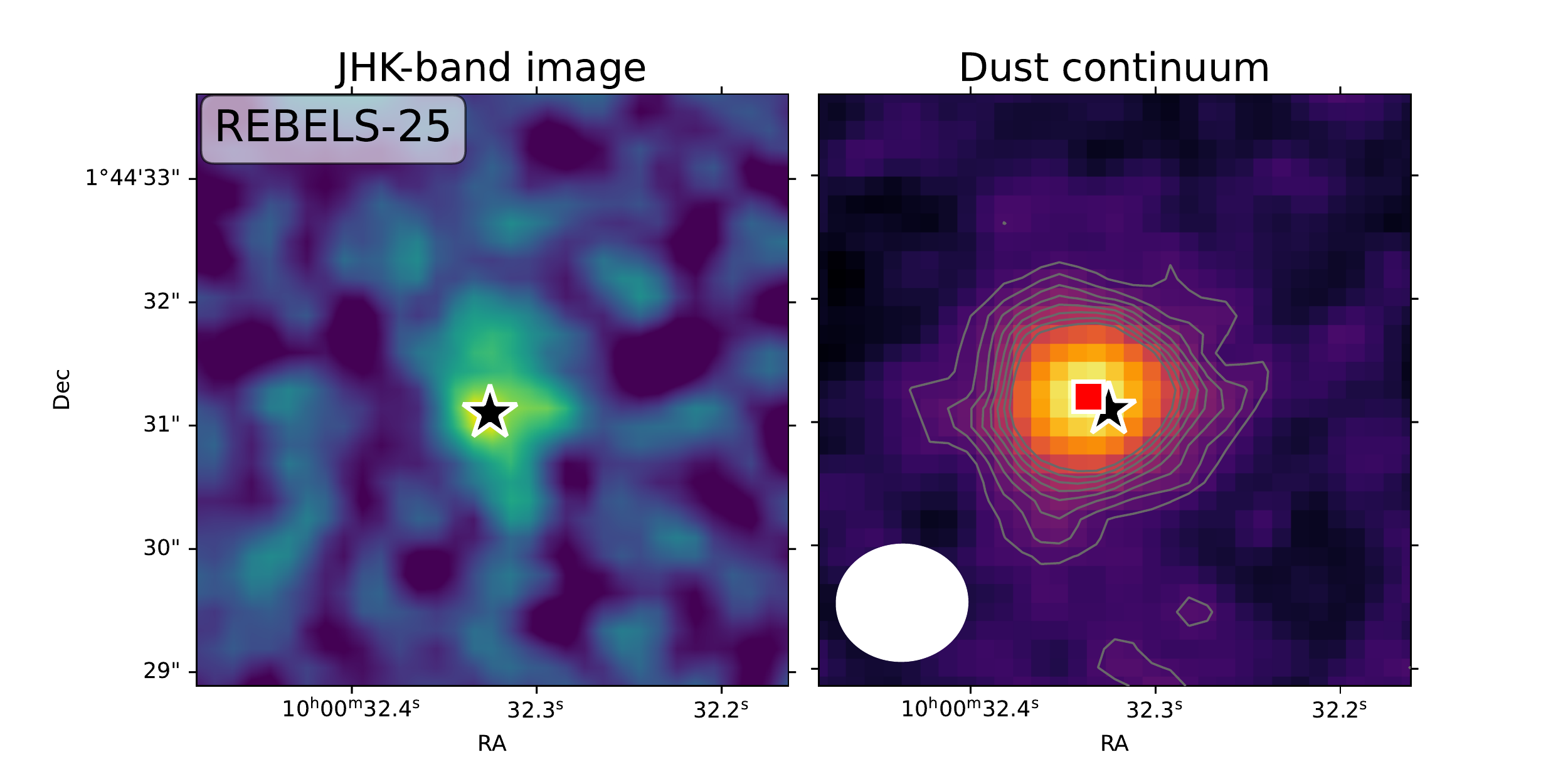}
    \includegraphics[scale=0.26, clip, trim=20 10 60 20]{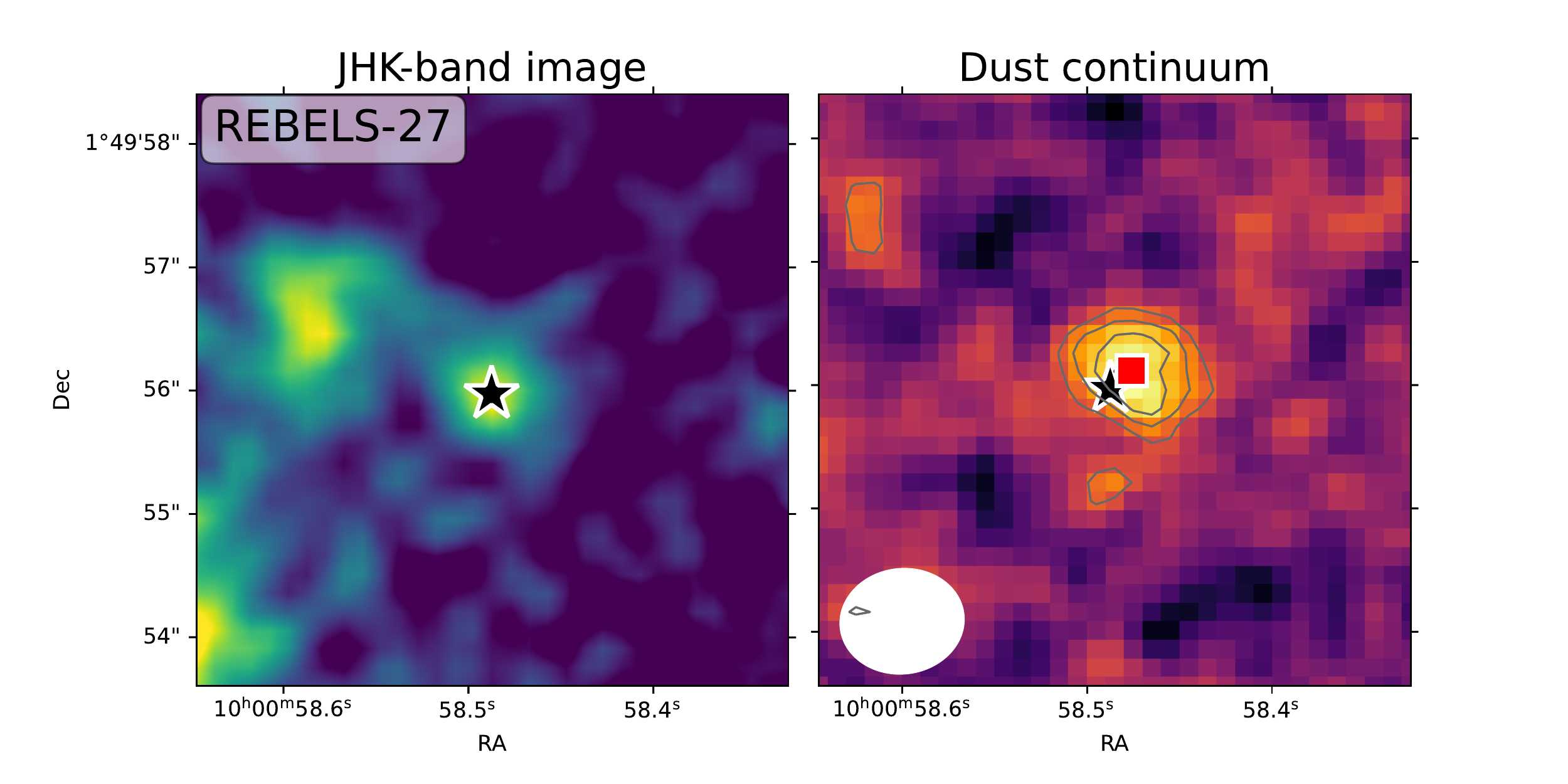}
    \includegraphics[scale=0.26, clip, trim=20 10 60 20]{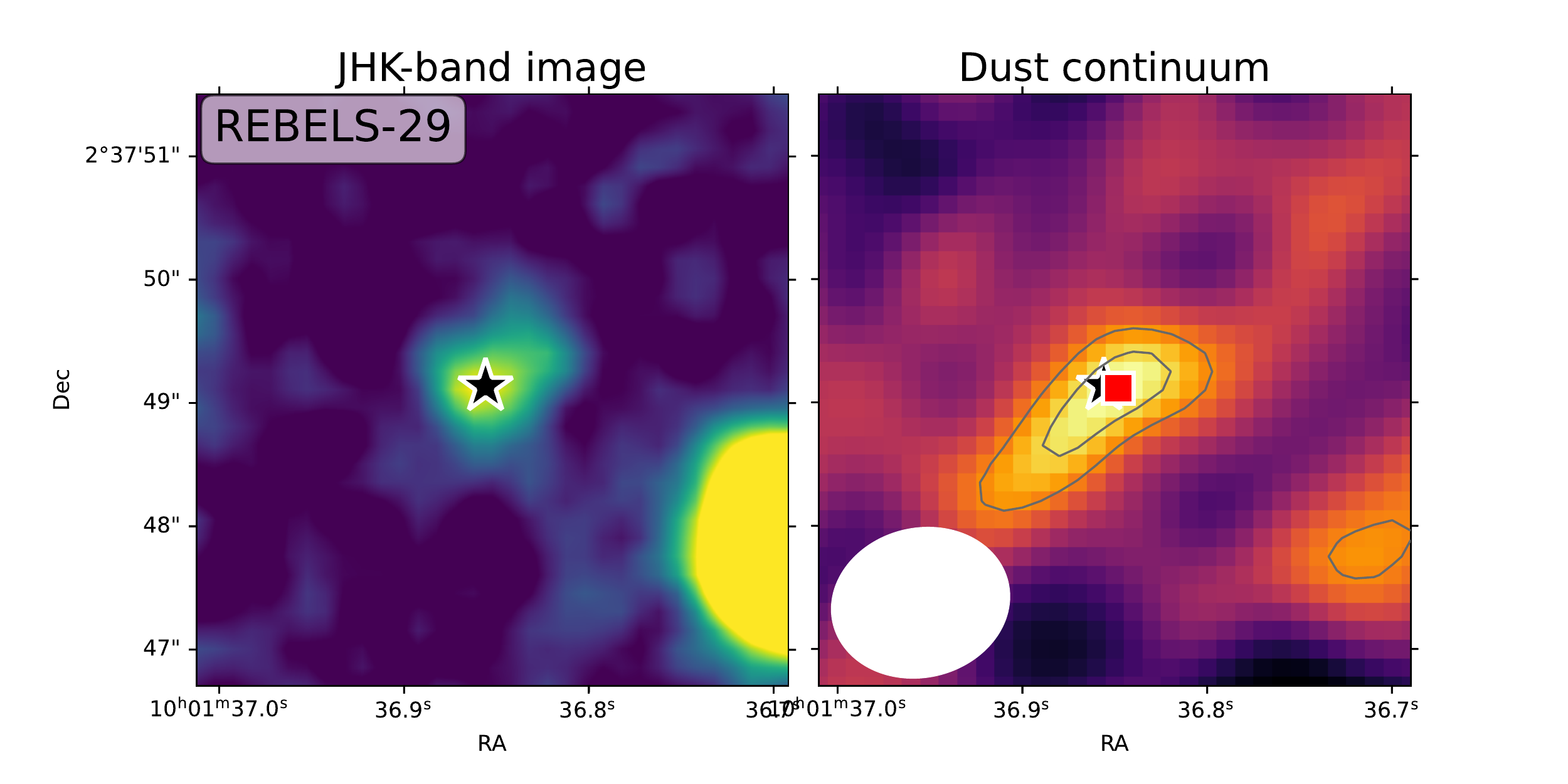}
    \includegraphics[scale=0.26, clip, trim=20 10 60 20]{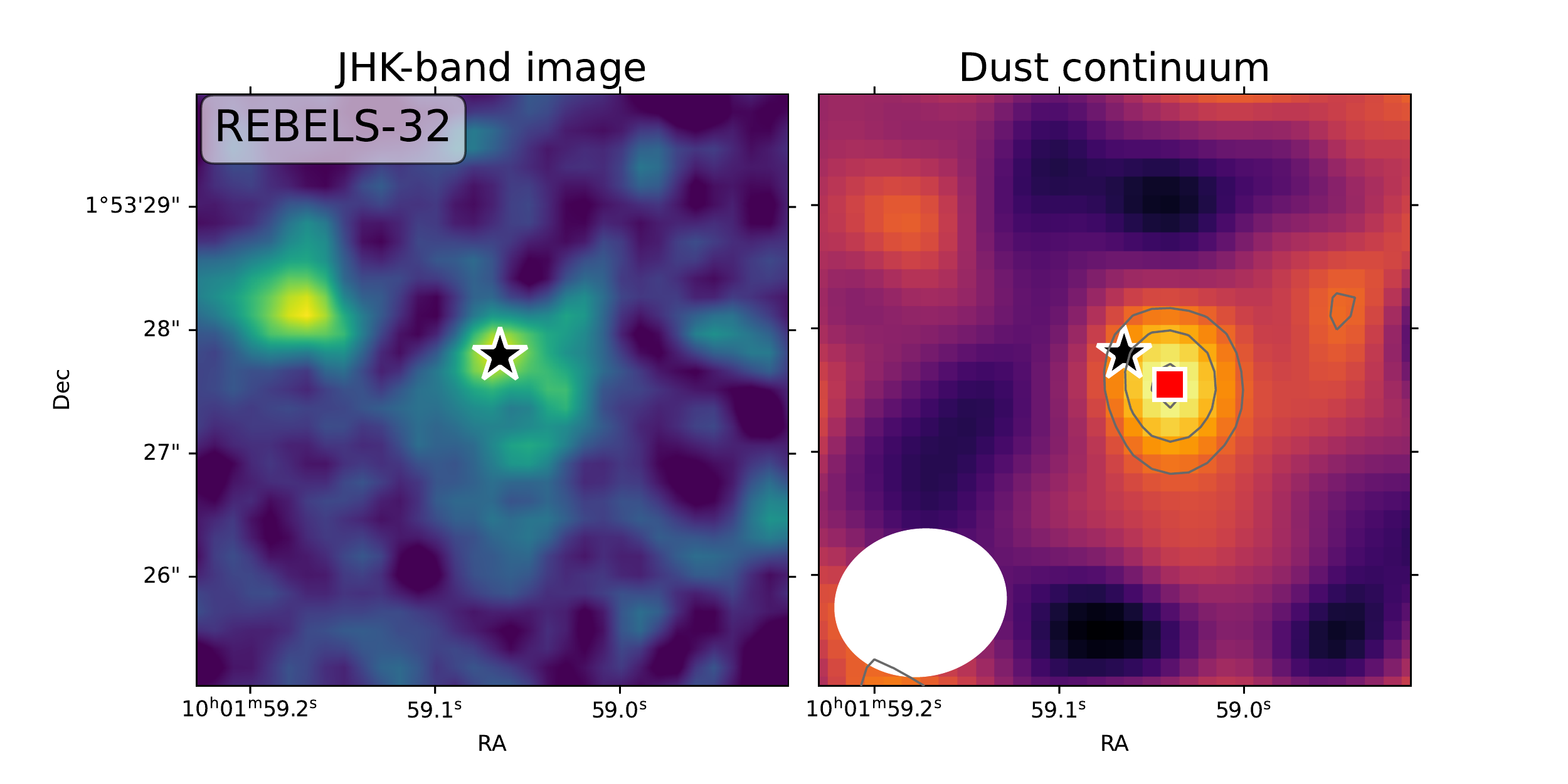}
    \includegraphics[scale=0.26, clip, trim=20 10 60 20]{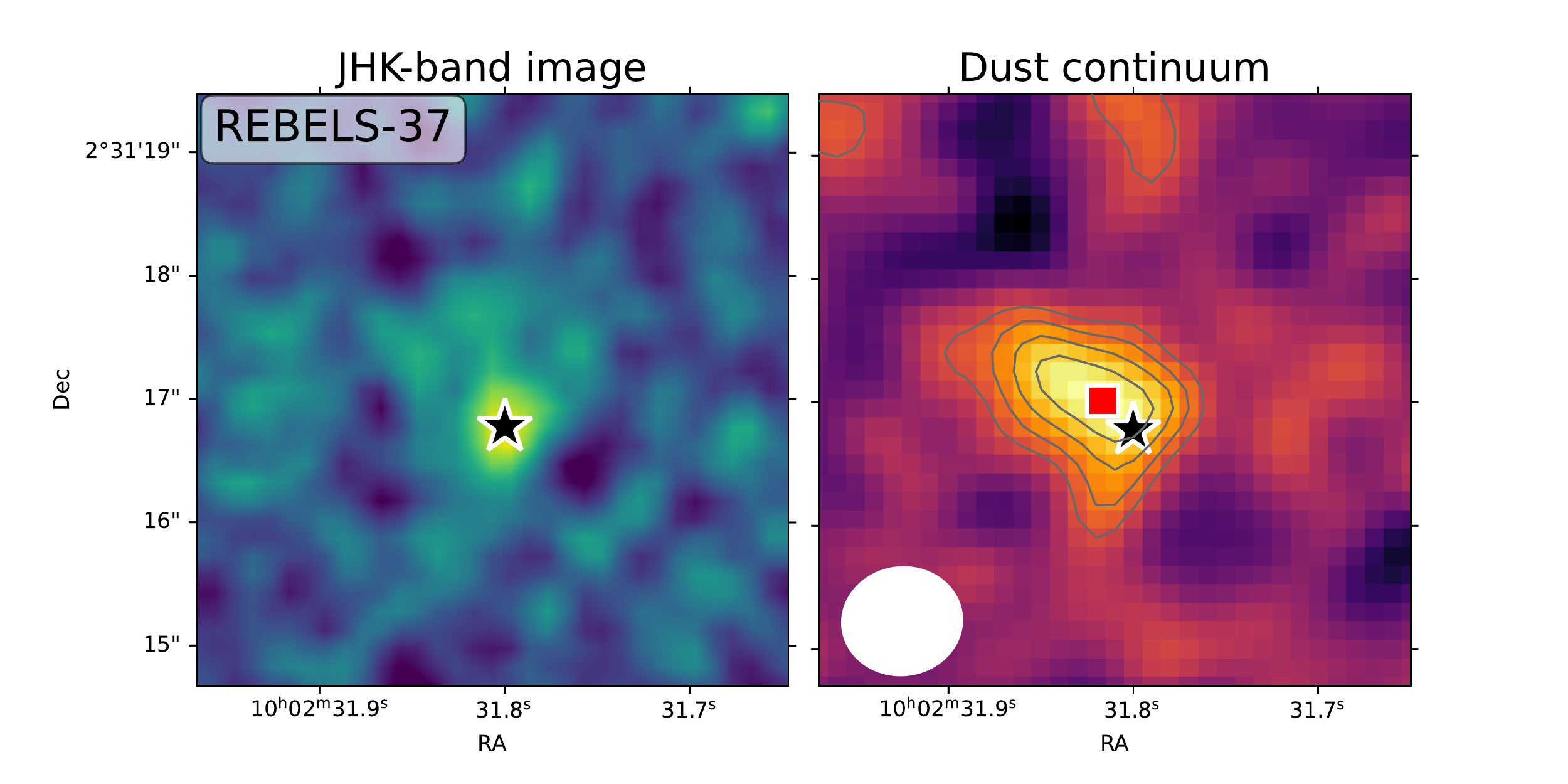}
    \includegraphics[scale=0.26, clip, trim=20 10 60 20]{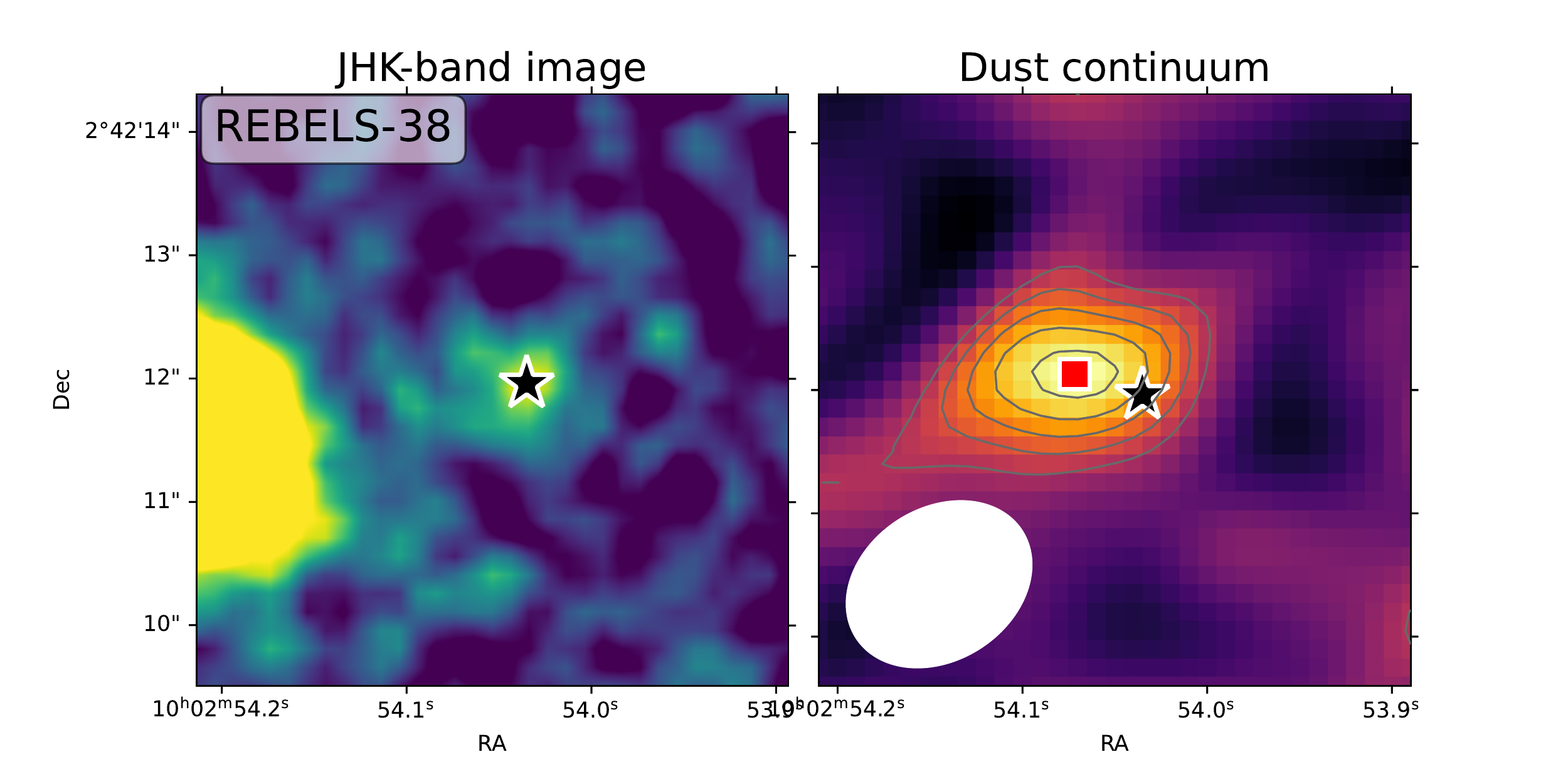}
    \includegraphics[scale=0.26, clip, trim=20 10 60 20]{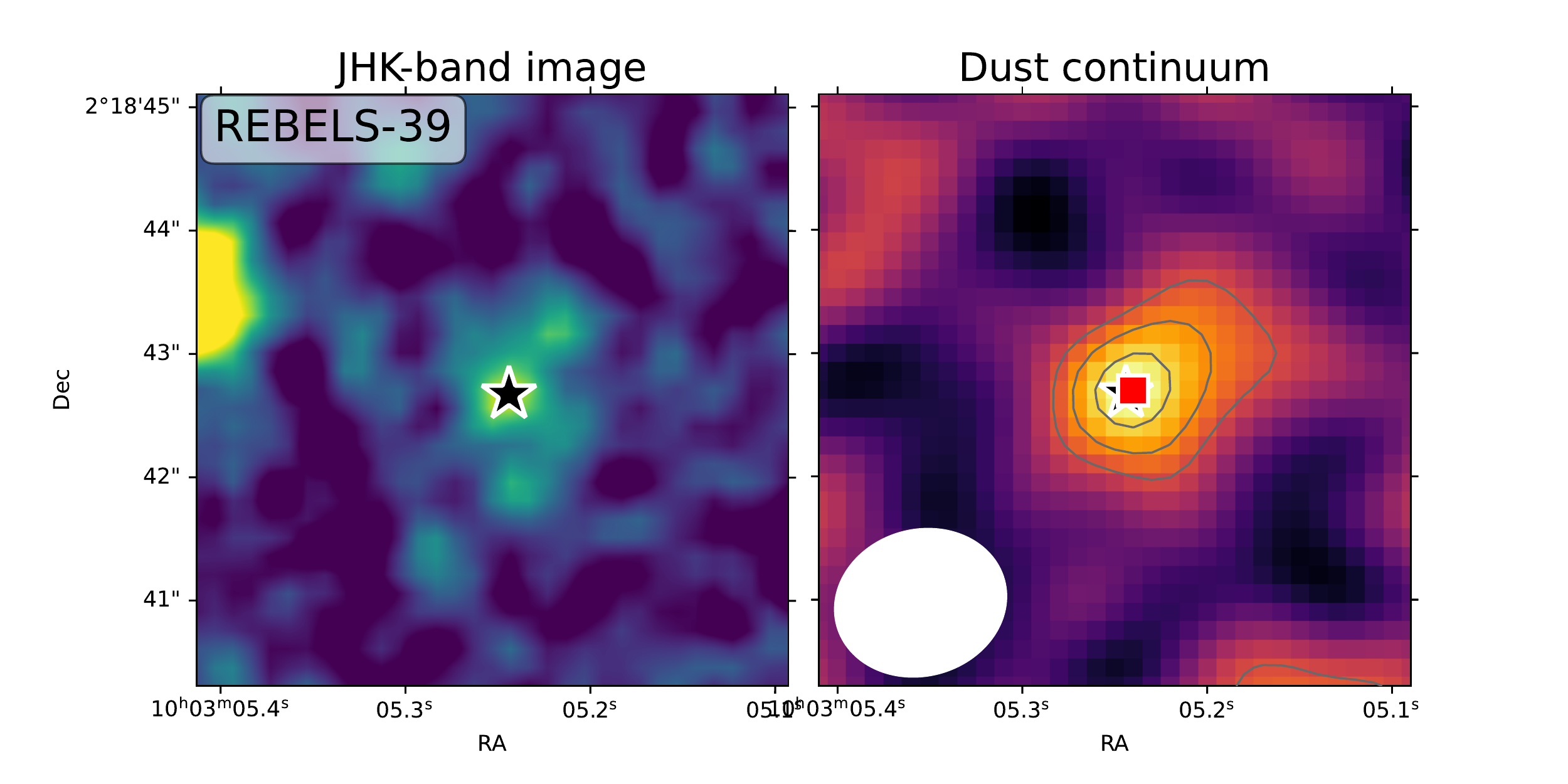}
    \includegraphics[scale=0.26, clip, trim=20 10 60 20]{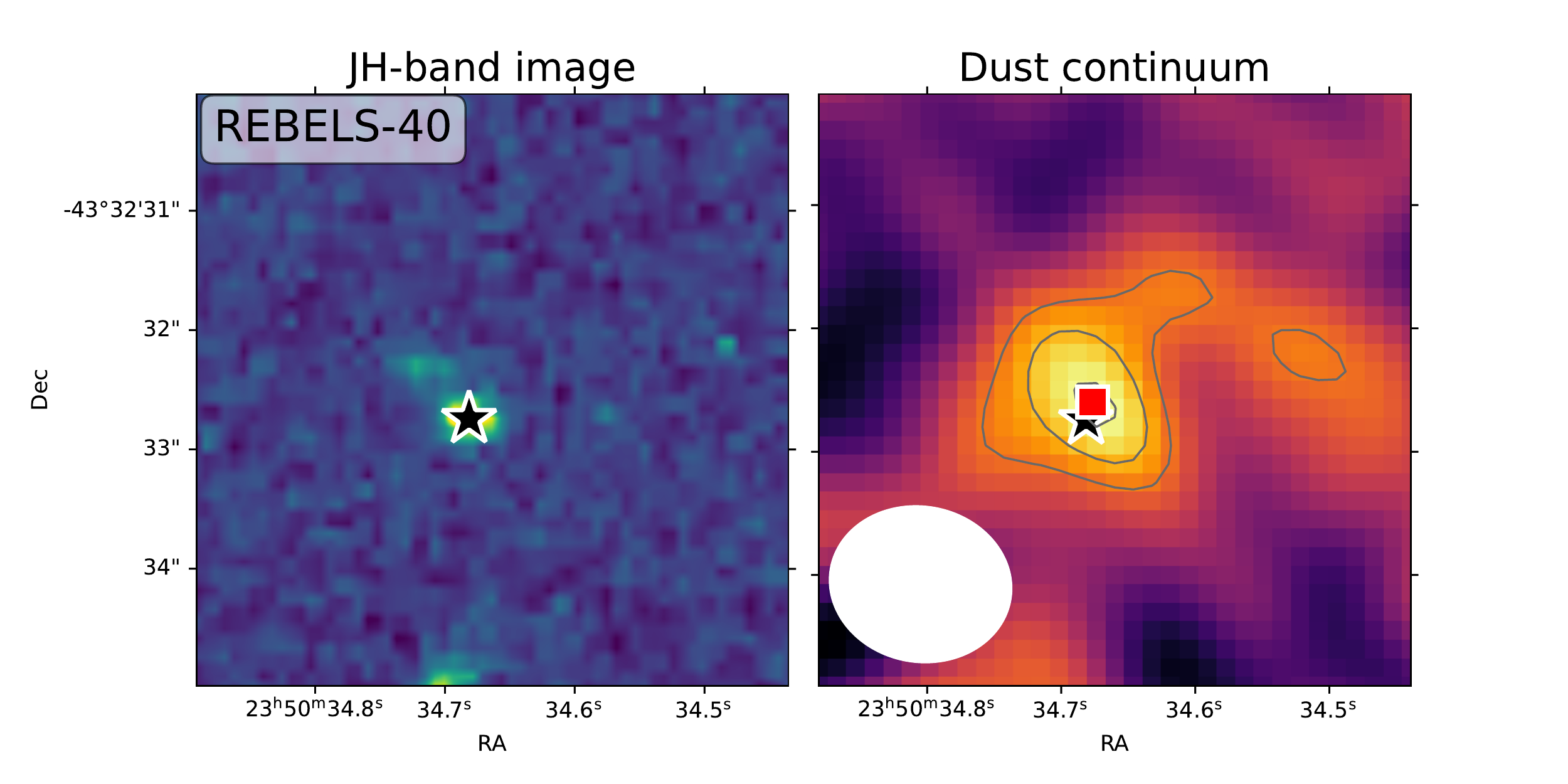}
    \includegraphics[scale=0.26, clip, trim=20 10 60 20]{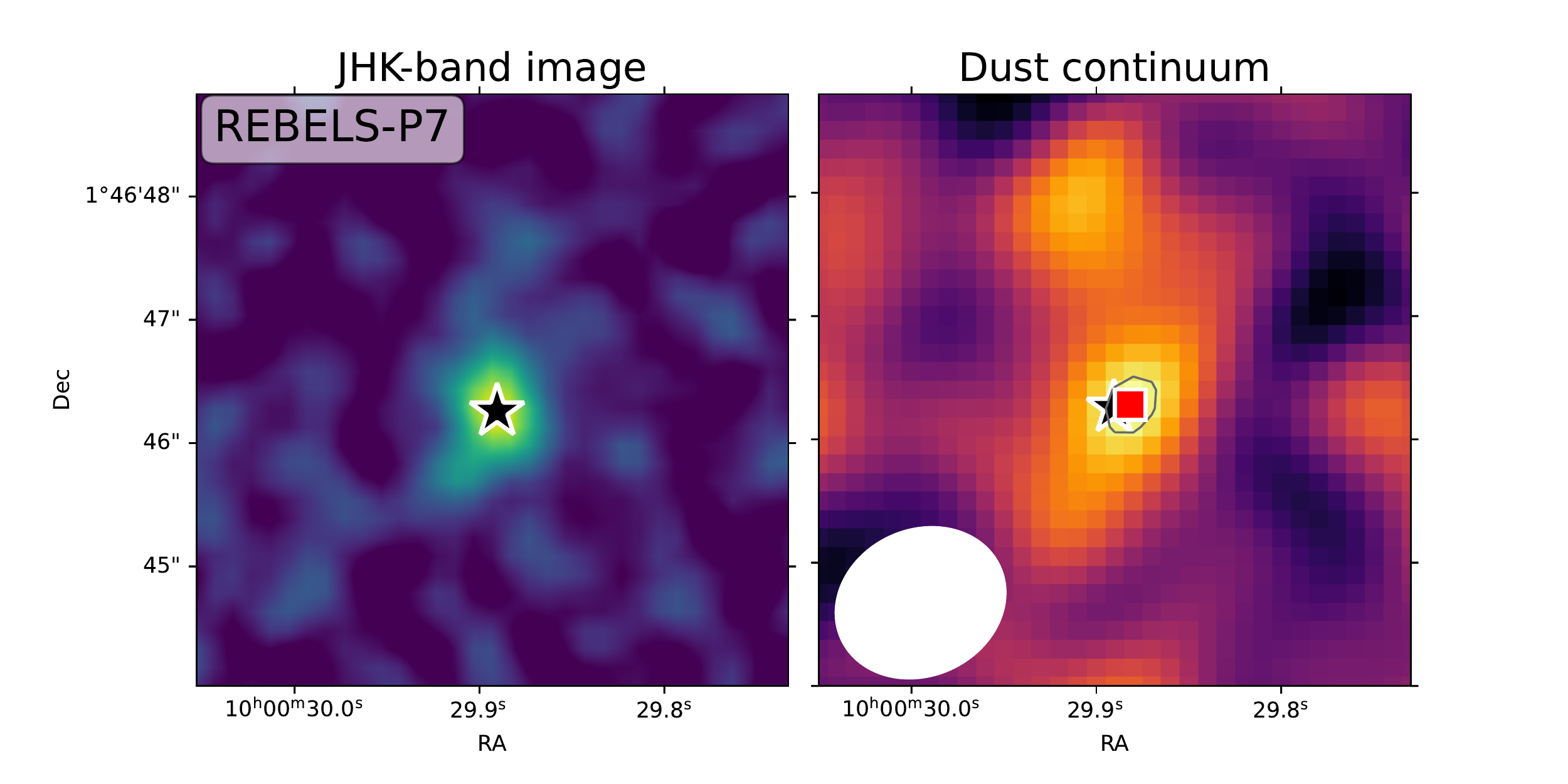}
    \includegraphics[scale=0.26, clip, trim=20 10 60 20]{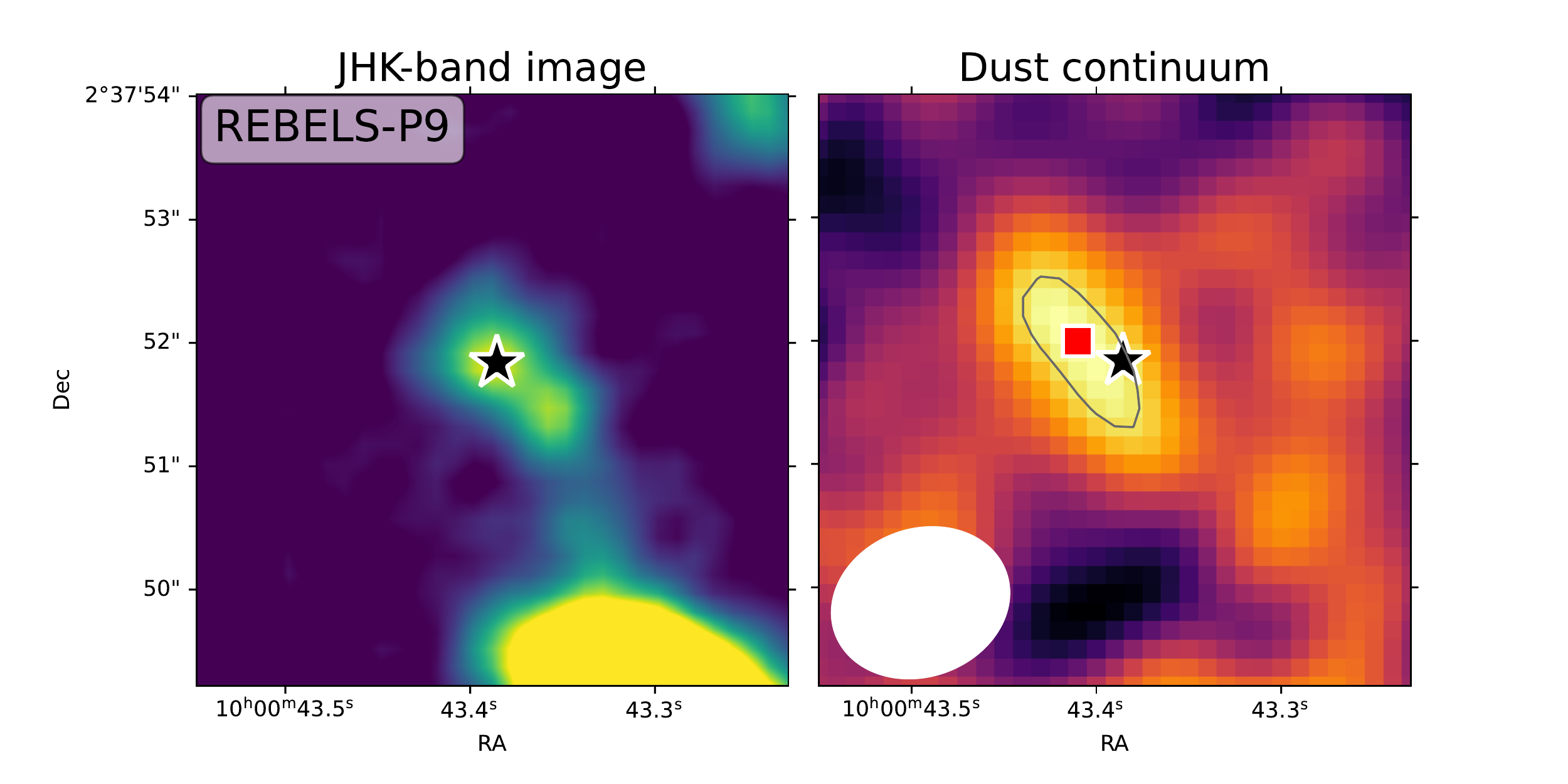}
    \caption{Peak locations of rest-UV (on the left of each panel) and
      dust (on the right of each panel) emission of the REBELS dust
      continuum detected sources. The peaks of the 
      UV and dust emission are indicated as the black star and red
      square, respectively. The rest-UV image shown here is a
      stack of $JHK$-band images (or $JH$ when $K$ is not available). 
      \label{fig:peak_pos}
    }
  \end{center}
\end{figure*}

\subsection{Infrared properties of the dust-detected galaxies at
  $\mathbf{z \sim 7}$}\label{subsec:IRprop}

All of the dust continuum detected galaxies have $L_{\rm IR} \gtrsim 3
\times 10^{11} \,{\rm L_{\odot}}$ and the detections are present up to 
$z \sim 8.5$ (Figure~\ref{fig:LIR}).  They are in the class of luminous
infrared galaxies (LIRGs) with $10^{11} < L_{\rm IR}/{\rm L_{\odot}} <
10^{12}$, except for REBELS-25 which is instead classified as an ultra
luminous infrared galaxy (ULIRG, $10^{12} < L_{\rm IR}/{\rm L_{\odot}} <
10^{13}$; \Hygate).  According to the model of \cite{Zava21}, the
obscured star formation is dominated ($\sim 80\%$) by galaxies with
$L_{\rm IR} > 10^{12}\,{\rm L_{\odot}}$ at $z\sim7$. Therefore, we might
have expected to detect more galaxies with $L_{\rm IR} > 10^{12}
\,{\rm L_{\odot}}$ in the REBELS sample.  This discrepancy could be due to the
REBELS target selection being based on rest-UV luminosity, which may
miss most obscured populations, as opposed to a blind survey.  
In addition, the discrepancy could arise from the
assumed faint-end slope of the infrared luminosity function in the
model of \cite{Zava21}. They adopted the faint-end slope from the
luminosity functions at $1 \lesssim z \lesssim 3$ derived by the
ASPECS program at 1mm and 3mm \citep{Gonz19,Gonz20,Arav20}. The 2mm
survey performed by \cite{Zava21} \citep[and ][]{Case21} covers a
wider but shallower area, and thus provides a robust constraint for
more luminous sources.  If there is an evolution in the faint-end
slope to become steeper towards the distant universe, then
galaxies with lower infrared luminosities would
contribute more to the obscured star formation at $z\sim7$.  An
infrared luminosity function based on the REBELS sources will be shown
in \Barrufet and the obscured star formation density at $z\sim7$ will
be presented in \Algera.

Despite the UV selection of the REBELS targets, the fraction of dust
obscured star formation, $\rm SFR_{IR}/SFR_{UV+IR}$, is high in the
dust continuum detected galaxies. As shown in
  Figure~\ref{fig:fobs}, with the REBELS detection limit, any galaxy
  with a dust continuum detection is $\gtrsim 50\%$
  obscured.  The obscured fraction ranges from $\approx 50\%$ to
$\approx 90\%$ \citep{Scho21,Bowl21b}. This is in
  agreement with the obscuration of $\sim 50-90\%$ 
  (except one source with 28\% but agrees within
  the errors) with an independent method in the REBELS sources 
  found by \cite{Ferr22} \citep[see also][]{Daya22}.  
  \Schneider have also investigated this wide spread of the obscuration 
  fraction with simulations in detail.  The obscured fraction of 
  our $z \sim 7$ galaxies is in a similar range to the ALPINE galaxies 
  at $z \sim 5$ with direct dust continuum detections \citep{Fuda20b}.

As shown in the left panel of Figure~\ref{fig:fobs}, the obscured 
fraction of our $z\sim7$ galaxies is on average lower 
for a given SFR, compared to the trend of
a stellar mass-completed galaxy sample at $z \lesssim 2.5$ 
\citep[][with the \cite{Dale02} SED templates]{Whit17}. 
The slope of the obscured fraction evolves from $z \sim 2.5$ 
to $z \sim 0$ \citep[e.g.,][]{Redd06,Redd08,Whit17}.  
Although the lower obscured fraction of the
REBELS galaxies could be due to the sample selection of 
UV bright galaxies (yet the obscured fraction is $\approx 50-90\%$), 
this may suggest that the evolution continues from $z \sim 7$.

On the contrary, the relation between obscured fraction and stellar
mass does not evolve between $0 \lesssim z \lesssim 2.5$ (the right
panel of Figure~\ref{fig:fobs}; see also Fig.~6 in
  \citealt{Daya22}).  Given that the stellar mass measurements of 
  the REBELS galaxies are still uncertain \citep[and could be underestimated 
  in the cases of constant SFH assumptions;][]{Topp22}, 
  the REBELS galaxies may be 
  in general consistent with the relation found at low redshift.
  There are a couple of outliers, such as
  REBELS-25 (the brightest continuum detection) and serendipitously
  detected sources without a detection in the rest-UV \citep{Fuda21},
  that lie above the $\rm SFR_{IR}/SFR_{UV+IR}-M_*$ relation.
  Excluding these extreme sources, the dependence of the obscured
  fraction on stellar mass at $0 \lesssim z \lesssim 2.5$ and $z \sim
  7$ may not evolve significantly.  Interestingly, however, the ALPINE
  dust detected galaxies at $z\sim5.5$, mostly lie below the relation
  at $0 \lesssim z \lesssim 2.5$ \citep{Fuda20b}. In the stellar mass
  range of $10^9-10^{10} \,{\rm M_\odot}$, where $2/3$ of the REBELS dust
  detected galaxies are, the stack of the ALPINE galaxies does not
  detect any dust continuum emission. This could be due to the
  difference in the sample selection between ALPINE and
  REBELS~\footnote{The ALPINE targets were selected
    based on UV spectroscopy and cover a lower stellar mass range than
    the REBELS sample.}, but may also imply a wide variety of
  fractions of obscured star formation at $z > 2.5$.
  More discussions on the obscured fraction at
    $z\sim7$, including a stacking analysis, will be presented in
    \Algera.  Better constraints on \LIR with ALMA and stellar mass
  with {\it JWST} are imperative to further
  investigate overall dust obscuration and its relationship to stellar
  mass and SFR of $z \sim 7$ galaxies.

\section{Dust Morphology and Spatial Offset between UV and IR} \label{sec:morph_offset}

We discern a small number of galaxies resolved in
  rest-frame UV imaging, while the majority of the dust emission is
marginally resolved or unresolved in our data with the spatial
resolution of $\sim 1.2-1.6\arcsec$.  In addition, we see physical
displacements between rest-UV and far-IR emission in a handful of the
REBELS dust continuum detected galaxies.

\subsection{Dust Morphology} \label{subsec:morph}

It is notable that two of the dust continuum detected targets,
REBELS-12 and REBELS-19, consist of two dust emission components. In
both cases, the component closer to where the UV emission is located
(the primary component, hereafter) is fainter than the other
component. Both of the galaxy systems also have a \CII detection, but
neither of them shows double peaks similar to the dust
emission. With the current data, it is difficult to assess whether
both of the dust components originate from the target galaxy or
whether one of them is from another unknown galaxy.

For REBELS-12, there is another rest-UV emitting source shown in the
$JHK$-stacked image with a $0.8\arcsec$ separation from the secondary
dust component (the dust emission component which is offset from the
UV location of the target source).
Interestingly, in the same FoV (primary beam) of the REBELS-12
observation, there is a serendipitous detection of a galaxy at the
same redshift confirmed by \CII~\footnote{The REBELS main target is at
$z=7.347$ and the serendipitous galaxy is at $z=7.352$
with a $\sim11.5\arcsec$ spatial separation.} with
dust continuum emission \citep{Fuda21}. It is not clear with the
current data if the offset dust component of REBELS-12 is related to
this overdensity.

In the case of REBELS-19, there is no obvious object in the
$JHK$-stacked image that may be associated with the secondary dust
component. Its \CII emission also shows an offset from the UV
emission, but peaking between the two dust components.  The
morphology of the UV emission seems to be elongated in the north-south
direction, along the dust emission.  Deeper observations and a higher
spatial resolution in both dust continuum and UV are necessary to
further investigate their physical structures.

Among galaxies with a single dust component, there is one object,
REBELS-25, that appears to be clearly resolved in both rest-UV and IR
(see \S\ref{subsec:flux_meas}).  This object has the highest \LIR in
the REBELS sample with $1.5^{+0.8}_{-0.5} \times 10^{12}\,{\rm L_{\odot}}$.
Its high S/N detection allows us to infer that this
  source is resolved at a $<1 \arcsec$ scale. The deconvolved size
reported by \texttt{imfit} is $(0.74\arcsec\pm0.17\arcsec) \times
(0.69\arcsec\pm0.22\arcsec)$, corresponding to $(3.75\pm0.89)\,{\rm
  kpc} \times (3.53\pm1.10) \,{\rm kpc}$.  This source also shows a
spatial offset towards the west in-between the clumpy UV emission
\citep{Stef19,Scho21}. For a more detailed study on REBELS-25, we
refer to \Hygate.

Although our Monte Carlo simulations (\S\ref{subsec:flux_meas}) do not
indicate that the other sources are resolved, there are some sources
that potentially show elongated emission. In fact, an observation with
deeper and higher spatial resolution conducted by \cite{Bowl21b} has
identified an extended tail in REBELS-29. In the REBELS data, we also
discern this tail but with less significance.

For the rest of the unresolved REBELS galaxies, unfortunately the
limited spatial resolution makes it hard to discuss their morphology
in further detail.  The spatial resolution of $\sim 1.2\arcsec$ of our
observations corresponds to a physical size of $\sim 6.3\,{\rm kpc}$
at $z=7$.  This is at the high end of the dust size measured at $z\sim
0-2$ \citep{Chen20, Tada20}.  Higher spatial resolution observations
are needed to investigate whether a more compact dust extent compared
with stellar emission is also seen at $z\sim7$ to explore mechanisms
of stellar mass growth in the first billion years of the universe
\citep[e.g.,][]{Hodg20, Ivis20, Inou20, Herr21, Popp21}.

\begin{figure}
  \begin{center}
    \includegraphics[width=0.45\textwidth]{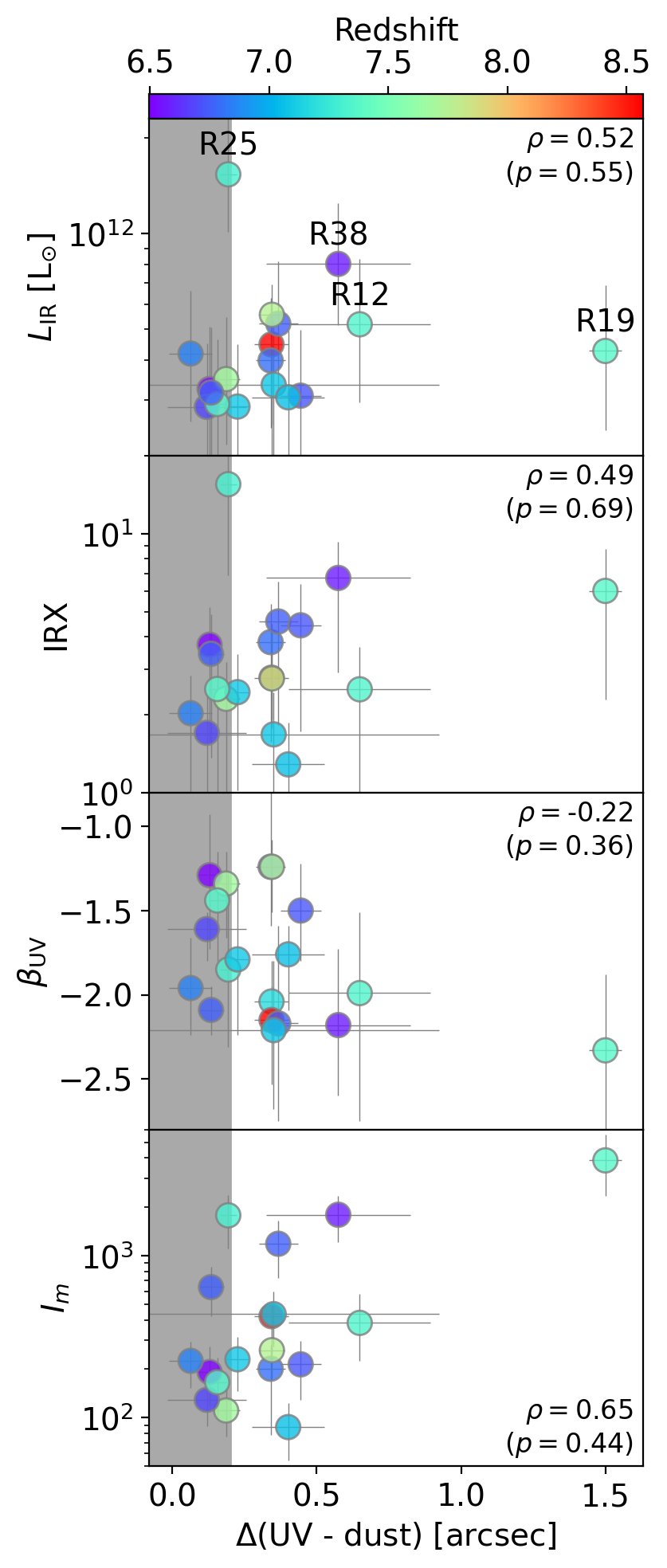}
    \caption{Galaxy physical properties against spatial offsets.  From
      the top to bottom, infrared luminosity (\LIR), UV spectral slope
      ($\beta_{\rm UV}$), infrared excess (IRX), and molecular index
      ($I_m$). The symbols are colour-coded by their spectroscopic
      redshifts where available, and the others are photometric
      redshifts. The grey region indicates the
        separation below the expected astrometric uncertainty of the
        ALMA observations (see \S\ref{subsec:dust_img}). 
        The Spearman rank correlation coefficient ($\rho$) and 
        the associated $p$-value are shown in each panel.
      \label{fig:offset}
    }
  \end{center}
\end{figure}

\begin{center}
  \begin{table}
\centering
\caption{Measured offset between the UV and dust peak emission, listed together with $\beta_{\rm UV}$ \citep{Bouw21b} and $I_m$ \citep{Ferr22}. For REBELS-04, REBELS-06 and REBELS-37, due to a lack of a spectroscopic redshift, as indicated by daggers, their photometric redshifts are used instead to calculate $I_m$.
\label{tbl:dust_offset}}
\begin{tabular}{cccc}
  \hline
  REBELS ID & $\Delta$(UV-dust) &          $\beta_{\rm UV}$ &                    $I_m$ \\ 
            &            arcsec &                           &                         \\ 
   \hline
  REBELS-04 & $ 0.34 \pm  0.06$ & $-2.15\,^{+0.20}_{-0.38}$ & $^{\dagger}$$   423\,^{+ 172}_{-  95}$  \\ 
  REBELS-05 & $ 0.13 \pm  0.04$ & $-1.29\,^{+0.36}_{-0.44}$ & $   191\,^{+  75}_{-  84}$  \\ 
  REBELS-06 & $ 0.34 \pm  0.05$ & $-1.24\,^{+0.67}_{-0.35}$ & $^{\dagger}$$   200\,^{+ 122}_{-  80}$  \\ 
  REBELS-08 & $ 0.37 \pm  0.07$ & $-2.17\,^{+0.58}_{-0.58}$ & $  1183\,^{+ 457}_{- 457}$  \\ 
  REBELS-12 & $ 0.65 \pm  0.25$ & $-1.99\,^{+0.48}_{-0.76}$ & $   384\,^{+ 161}_{- 197}$  \\ 
  REBELS-14 & $ 0.35 \pm  0.57$ & $-2.21\,^{+0.41}_{-0.47}$ & $   435\,^{+ 159}_{- 166}$  \\ 
  REBELS-18 & $ 0.19 \pm  0.05$ & $-1.34\,^{+0.19}_{-0.32}$ & $   111\,^{+  34}_{-  40}$  \\ 
  REBELS-19 & $ 1.50 \pm  0.06$ & $-2.33\,^{+0.45}_{-0.64}$ & $  3870\,^{+1544}_{-1719}$  \\ 
  REBELS-25 & $ 0.19 \pm  0.03$ & $-1.85\,^{+0.56}_{-0.46}$ & $  1772\,^{+ 660}_{- 585}$  \\ 
  REBELS-27 & $ 0.23 \pm  0.04$ & $-1.79\,^{+0.42}_{-0.45}$ & $   229\,^{+  84}_{-  86}$  \\ 
  REBELS-29 & $ 0.12 \pm  0.14$ & $-1.61\,^{+0.10}_{-0.19}$ & $   129\,^{+  40}_{-  42}$  \\ 
  REBELS-32 & $ 0.44 \pm  0.07$ & $-1.50\,^{+0.28}_{-0.30}$ & $   213\,^{+  84}_{-  85}$  \\ 
  REBELS-37 & $ 0.34 \pm  0.04$ & $-1.24\,^{+0.16}_{-0.27}$ & $^{\dagger}$$   261\,^{+  89}_{- 133}$  \\ 
  REBELS-38 & $ 0.57 \pm  0.25$ & $-2.18\,^{+0.45}_{-0.42}$ & $  1786\,^{+ 571}_{- 555}$  \\ 
  REBELS-39 & $ 0.06 \pm  0.08$ & $-1.96\,^{+0.30}_{-0.28}$ & $   224\,^{+  72}_{-  71}$  \\ 
  REBELS-40 & $ 0.16 \pm  0.03$ & $-1.44\,^{+0.29}_{-0.36}$ & $   165\,^{+  64}_{-  69}$  \\ 
  REBELS-P7 & $ 0.13 \pm  0.04$ & $-2.09\,^{+0.14}_{-0.15}$ & $   639\,^{+ 215}_{- 217}$  \\ 
  REBELS-P9 & $ 0.40 \pm  0.12$ & $-1.76\,^{+0.17}_{-0.33}$ & $    87\,^{+  33}_{-  36}$  \\ 
  \hline
\end{tabular}
\end{table}

\end{center}

\subsection{Spatial Offset between rest-UV and IR}\label{subsec:offset}

Spatial offsets in the peak emission between the rest-frame UV and
far-infrared have become evident in galaxies at $z > 6.5$ as well as
at lower redshifts with recent high spatial resolution observations at
submillimetre and millimetre wavelengths (see e.g.,
\citealt{Hodg12,Hodg16,Carn17,Lapo17a,Rujo19} for observational, and
e.g., \citealt{Behr18, Lian19, Somm20, Coch21} for theoretical
studies).  The spatial displacement between the rest-UV and far-IR
could depend on the process of how dust and stars have been assembled
\citep[e.g.,][]{Coch19,Zane21,Ferr22}, but also bias analyses that are based
on the UV-IR energy balance \citep[e.g., IRX-$\beta$, SED
  fitting,][]{Bari17,Bowl21b}. Here, we explore the prevalence of a
spatial offset between the rest-UV and dust emission in UV-selected
 galaxies at $z\sim7$ from REBELS and how it relates to the
galaxy properties.

We determine the peaks of the rest-UV and dust
continuum emission by a 2D Gaussian fit to avoid being biased to a
  clump or accidentally picking up a noise peak. We let the
amplitude, xy positions, and rotation be free parameters for the fit,
whereas the xy width and background level are fixed to 1 pixel and
zero, respectively. We show the determined peak
  locations of the rest-UV and dust emission in
  Figure~\ref{fig:peak_pos}. 

Based on the defined peak locations, we then calculate the spatial
offsets between the rest-UV and dust emission. 
The measured offsets are listed in Table~\ref{tbl:dust_offset}. 
The two galaxies, REBELS-12 and 19, with the double dust emission 
peaks show the largest offsets among the sample, $0.7\arcsec$ (3.5\,kpc) 
and $1.5\arcsec$ (7.6\,kpc), respectively (see also \S\ref{subsec:morph} 
for their morphologies). REBELS-38 also has a prominent offset of 
$0.6\arcsec$ (3.3\,kpc). The spatial offset of REBELS-25 has been 
reported in \cite{Scho21} and is discussed in more detail in \Hygate.

To explore whether the UV-IR spatial segregation affects other observed
physical properties of the galaxies, we compare the UV-dust offset to
galaxy properties in Figure~\ref{fig:offset}.  
The top three panels show comparisons of \LIR, IRX, 
and $\beta_{\rm UV}$ against the the UV-IR separation, but no
significant trend has been found.
If a correlation with IRX or the UV slope exists,
galaxies being outliers in the IRX-$\beta_{\rm UV}$ relation could 
be due to a larger spatial separation between the rest-UV
and dust emission peaks. This may be related to the recent findings
of \cite{Bowl21b}, who presented the IRX-$\beta_{\rm UV}$ relation of
five galaxies at $z\sim7$ with UV-dust spatial offsets. All of these 
galaxies (one is reported in this work as well) have $\beta_{\rm UV}
\lesssim -2.0$. It is possible that galaxies with spatially decoupled
UV and IR emission tend to show an unusually blue colour in the UV (see
also theoretical work by \citealt{Behr18} and \citealt{Somm20}).
However, a larger sample of galaxies with a moderate to large spatial
offset observed with higher spatial resolution imaging 
is needed to confirm whether $\beta_{\rm UV}$ and centroid
locations of the UV and dust are physically related.

Motivated to search for this decoupled UV and IR emission,
we also compare the spatial offsets with $I_m$, the molecular index, 
defined as \[ I_m=\frac{F_{158}/F_{1500}}{\beta_{\rm UV}-\beta_{\rm UV, int}} \]
where $F_ {158}$ and $F_{1500}$ are the observed
  continuum flux densities at rest-frame 158\um and 1500\AA,
  respectively, and $\beta_{\rm UV, int}$ is the intrinsic UV slope
  from a model (see \citealt{Ferr22} for more details, as well as
\citealt{Pall17} and \citealt{Behr18}). 
A high value of $I_m$ implies a large IR-to-UV flux ratio (i.e., red
colour) compared to the galaxy UV spectral slope. Such values can only
be obtained if the galaxy has a multi-phase ISM made of star forming,
opaque clumps, and a diffuse component that is relatively transparent
to UV light emitted by young stars (see \citealt{Ferr22} for a more detailed
discussion). 
No statistically significant correlation is seen between
$I_m$ and the UV-IR separation, given the Spearman rank correlation 
coefficient is 0.65 with a $p$-value of 0.44.
Although we cannot exclude the null-hypothesis of no correlation, 
increasing $I_m$ with the UV-IR offset, in a simple picture,
indicates that star formation in these early galaxies has
a (possibly slightly older) star formation site where the dust has 
already been cleared up, accompanied by another younger star formation 
site still embedded in dusty molecular clouds.

There are indeed some galaxies in the epoch of reionisation that have
been speculated to contain two stellar populations to elucidate their
observed features.  Although it is not detected with dust emission,
the red rest-frame optical colour of a gravitationally lensed
star-forming galaxy MACS1149-JD1 at $z=9.1$ indicates that it already
experienced a star forming phase at $z \approx 15$ \citep[][also see
  the discussion in \citealt{Robe20}]{Hash18}.  Another example is
GN-z11, which first was discovered via a spectroscopically identified
rest-UV continuum break at $z=11$ \citep{Oesc16}. Recently, this
galaxy has been confirmed by \cite{Jian21} at this redshift with the
[\ion{C}{iii}]$\lambda 1907$, \ion{C}{iii}]$\lambda 1909$ doublet and
\ion{O}{iii}]$\lambda 1666$.  Assuming no contribution from an active
galactic nucleus, its observed high rest-frame equivalent width
($> 20$\AA) of \ion{C}{iii}] indicates the presence of a young
stellar population. Its moderate stellar mass,
$(1.3\pm0.6)\times 10^9 \,{\rm M_\odot}$, also requires a relatively 
evolved population to be present.  Additionally, the two $z\sim13$ 
candidates of star-forming galaxies reported by \cite{Hari21} may also 
support an onset of star formation at $z > 10$.
\cite{Manc16} found that in order to simultaneously reproduce the
observed UV luminosity function and the relation between $\beta_{\rm
  UV}$ and UV magnitude in their model, they have to assume a
two-phase distribution for dust, with two different dust optical depths.  
However, alternatively, it is also possible that a large $I_m$ is
caused by a merging system of an obscured galaxy and an
optically thin galaxy (cf. REBELS-25, \Hygate).  Although we
cannot draw firm conclusions regarding the mass and dust
assembly of these very early galaxies with the limited
resolution of the current data, our systematic observations of
dusty galaxies at $z\sim7$ provide important clues of the
build-up of stellar mass and dust in galaxies in the early
universe \citep[see also][]{Tacc21,Pall22,Daya22}.

\section{Conclusions}

In this paper, we presented the dust continuum source identifications
and flux extractions of the UV-selected galaxies at $z > 6.5$
of the ALMA Large Program REBELS and its pilot
programs. The main results are summarised as follows:

\begin{itemize}

\item Out of the 40 REBELS targets, we detected \NDetCont galaxies
  that have rest-frame $\sim88$\um or 158\um dust continuum emission
  with $\geq 3.3\sigma$ where the purity is 95\%.  There are still six
  targets whose observations remain to be completed, making the
  current detection rate $\geq 40\%$.  Together with the REBELS pilot
  programs, which adds an additional nine galaxies, we obtained 18
  dust continuum detections at $z > 6.5$ out of a
    total of 49 targets. This, in turn, increases the sample of dusty
    star-forming galaxies in the epoch of reionisation by a factor of
    more than three.

\item The spatial resolution of the observations ranged between
  $1.2-1.6\arcsec$. Based on Monte Carlo simulations, to identify
  resolved dust emission, we found that one galaxy (REBELS-25) shows a
  spatially resolved structure. In addition, two galaxies (REBELS-12
  and 19) show double components in their dust continuum emission.

\item Infrared luminosities of the dust continuum 
  detected galaxies are in a range of $3 \times
  10^{11} \lesssim L_{\rm IR}/{\rm L_\odot} \lesssim 2 \times 10^{12}$ 
  with one galaxy classified as a ULIRG.  Despite being
  UV-selected targets, the dust continuum detected galaxies have a
  high fraction of obscured star formation ($\sim 50-90\%$).

\item We also found that some of the dust continuum detected galaxies
  exhibit spatially decoupled rest-UV and far-IR emission. 
  However, with current limited spatial resolution imaging, 
  no clear trend has been seen in $L_{\rm IR}$, IRX, $\beta_{\rm UV}$, 
  and $I_m$ against the UV-IR separation. To confirm the speculated two 
  different populations in single galaxies (or a merging system) 
  during the epoch of reionisation, higher spatial resolution observations 
  are needed.

\end{itemize}
  
The REBELS program has provided a first statistical glimpse of
obscured star formation in UV-selected galaxies at $z > 6.5$.
In the near future, multi-band ALMA coverage will help to better
constrain dust SEDs and improve measurements on dust
  temperature and mass.  Together with resolved studies of individual
sources, this will facilitate our understanding of star formation and
dust buildup in the first billion years of the universe.

\section*{Acknowledgements}

The authors would like to thank the referee whose constructive 
comments helped improve the manuscript. 
We acknowledge assistance from Allegro, the European ALMA Regional
Center node in the Netherlands.
ALMA is a partnership of ESO (representing its member states), NSF
(USA) and NINS (Japan), together with NRC (Canada), NSC and ASIAA
(Taiwan), and KASI (Republic of Korea), in cooperation with the
Republic of Chile. The Joint ALMA Observatory is operated by ESO,
AUI/NRAO and NAOJ. This paper makes use of the following ALMA data:
ADS/JAO.ALMA\#2019.1.01634.L,
ADS/JAO.ALMA\#2017.1.01217.S,
ADS/JAO.ALMA\#2017.1.00604.S,
ADS/JAO.ALMA\#2018.1.00236.S,
ADS/JAO.ALMA\#2018.1.00085.S
ADS/JAO.ALMA\#2018.A.00022.S.
This work was supported by NAOJ ALMA Scientific Research Grant Code
2021-19A (HI and HSBA).
HI acknowledges support from JSPS KAKENHI Grant Number JP19K23462.
SS acknowledges support from the Nederlandse Onderzoekschool voor
Astronomie (NOVA).
RJB and MS acknowledge support from TOP grant TOP1.16.057.
RS and RAB acknowledge support from STFC Ernest Rutherford Fellowships
[grant numbers ST/S004831/1 and ST/T003596/1].
RE acknowledges funding from JWST/NIRCam contract
to the University of Arizona, NAS5-02015.
PAO, LB, and YF acknowledge support from the Swiss National Science
Foundation through the SNSF Professorship grant 190079 `Galaxy
Build-up at Cosmic Dawn'.
AF, AP and LS acknowledges support from the ERC Advanced Grant INTERSTELLAR
H2020/740120.
Generous support from the Carl Friedrich von Siemens-Forschungspreis
der Alexander von Humboldt-Stiftung Research Award is kindly
acknowledged (AF).
MA acknowledges support from FONDECYT grant 1211951, CONICYT + PCI +
INSTITUTO MAX PLANCK DE ASTRONOMIA MPG190030, CONICYT+PCI+REDES 190194
and ANID BASAL project FB210003.
PD acknowledges support from the European Research
Council's starting grant ERC StG-717001 (``DELPHI"), from the NWO
grant 016.VIDI.189.162 (``ODIN") and the European Commission's and
University of Groningen's CO-FUND Rosalind Franklin program.
JH gratefully acknowledges support of the VIDI research program with
project number 639.042.611, which is (partly) financed by the
Netherlands Organisation for Scientific Research (NWO).
LG and RS acknowledge support from the Amaldi Research Center funded
by the MIUR program ``Dipartimento di Eccellenza''
(CUP:B81I18001170001).
YF further acknowledges support from NAOJ ALMA Scientific Research
Grant number 2020-16B ``ALMA HzFINEST: High-z Far-Infrared Nebular
Emission STudies''.
IDL acknowledges support from ERC starting grant 851622 DustOrigin.
This work has made use of data from the European Space Agency (ESA)
mission {\it Gaia} (\url{https://www.cosmos.esa.int/gaia}), processed
by the {\it Gaia} Data Processing and Analysis Consortium (DPAC,
\url{https://www.cosmos.esa.int/web/gaia/dpac/consortium}). Funding
for the DPAC has been provided by national institutions, in particular
the institutions participating in the {\it Gaia} Multilateral
Agreement.
This paper utilizes observations obtained with the NASA/ESA Hubble
Space Telescope, retrieved from the Mikulski Archive for Space
Telescopes (MAST) at the Space Telescope Science Institute
(STScI). STScI is operated by the Association of Universities for
Research in Astronomy, Inc. under NASA contract NAS 5-26555.
This work is based (in part) on observations made with the Spitzer
Space Telescope, which was operated by the Jet Propulsion Laboratory,
California Institute of Technology under a contract with NASA. Support
for this work was provided by NASA through an award issued by
JPL/Caltech.

\section*{Data Availability}

Table~\ref{tbl:det_src} in this article is available at the 
Centre de Donn\'ees astronomiques de Strasbourg (CDS), at \url{http://vizier.u-strasbg.fr/viz-bin/VizieR?-source=J/MNRAS/515/3126}. 
The other data generated in this research will be shared on reasonable 
request to the corresponding author.



\bibliographystyle{mnras}
\bibliography{bib} 




\appendix

\section{NIR images and dust continuum contours of the entire sample}

\begin{figure*}
  \includegraphics[width=\textwidth, clip, trim=0 0 0 80]{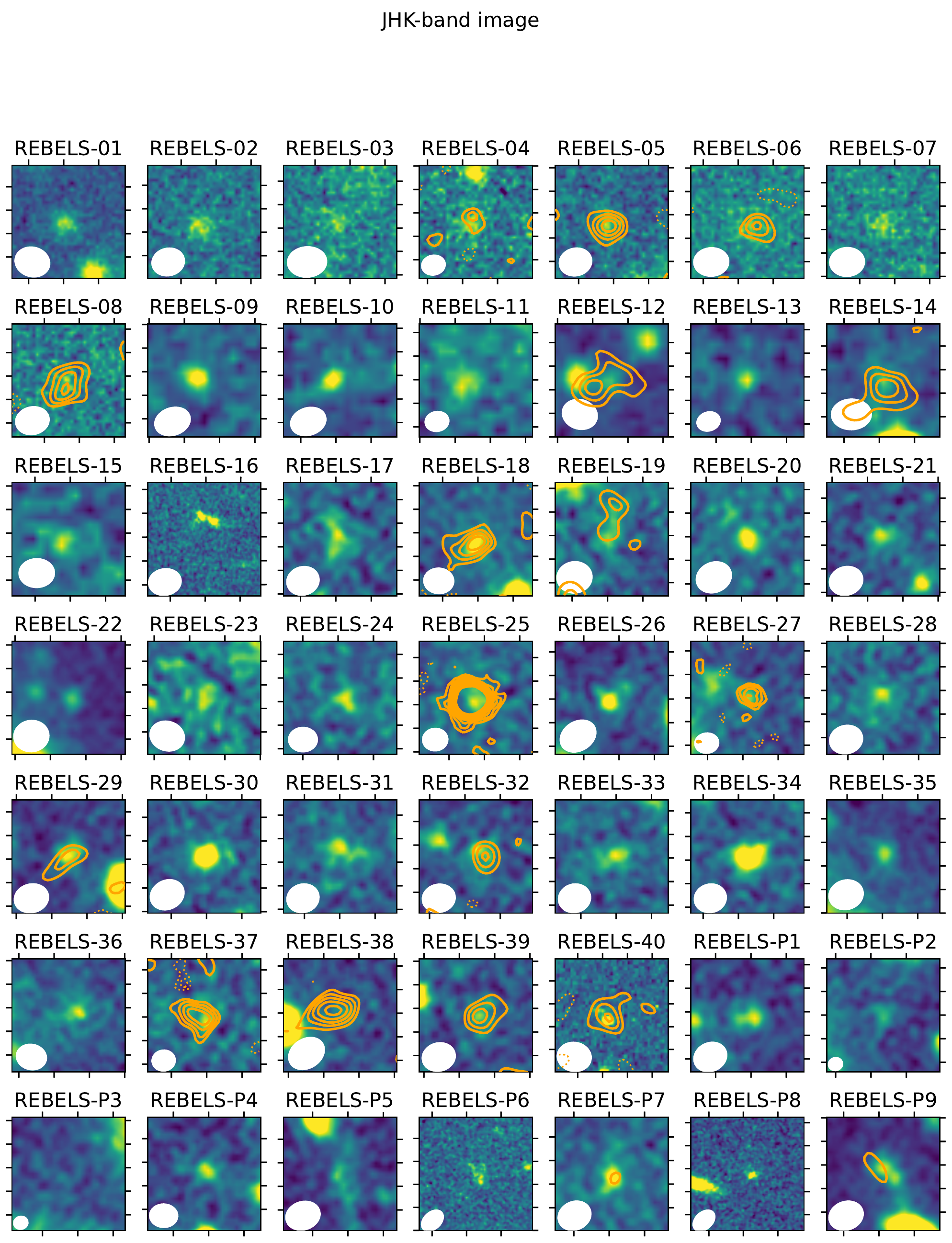}
  \caption{ The same as Figure~\ref{fig:dust_det} without the \CII
    contours. The targets without a dust continuum detection are also
    shown.
    \label{fig:dust_det_all}
  }
\end{figure*}

In Figure~\ref{fig:dust_det_all}, we show the stacked $JHK$-band images 
(or $JH$-band images when $K$-band image is not available)
with the contours of dust continuum emission for all of the REBELS
sample.

\bsp	
\label{lastpage}
\end{document}